\documentclass[acmlarge, nonacm]{acmart}
\usepackage{geometry}
\usepackage{graphicx} % Required for inserting images
\usepackage{hyperref}
\usepackage{adjustbox}
\usepackage{xcolor}
\usepackage[dvipsnames]{xcolor}
\usepackage{colortbl}
\usepackage{multirow}
\usepackage{enumitem}
\usepackage{booktabs}
\usepackage{comment}
\usepackage{wrapfig}
\usepackage{longtable}
\usepackage{ragged2e}
\usepackage{array}
\usepackage[table]{xcolor}
\usepackage[T1]{fontenc}
\usepackage[utf8]{inputenc}

\def\eg{\emph{e.g., }}

\newcommand{\am}[1]{\textcolor{black}{#1}}

\AtBeginDocument{%
  }

\setcopyright{acmlicensed}
\copyrightyear{2018}
\acmYear{2018}
\acmDOI{XXXXXXX.XXXXXXX}
\acmConference[Conference acronym 'XX]{Make sure to enter the correct
  conference title from your rights confirmation email}{June 03--05,
  2018}{Woodstock, NY}
\acmISBN{978-1-4503-XXXX-X/2018/06}

\begin{document}

%\title{Towards Collective Knowledge: Fragmentation in Evaluation of Ambient AI Scribes in Healthcare}
\title{Ninety-Six Ways to Measure a Scribe: Fragmented Evaluation of Ambient  Scribes in Healthcare}
\author{Amama Mahmood}
 \email{amama.mahmood@jhu.edu}
 \affiliation{%
   \institution{Johns Hopkins University}
  \streetaddress{3400 N. Charles St}
   \city{Baltimore}
   \state{Maryland}
   \country{USA}
   \postcode{21218}
 }

 \author{Shiye Cao}
 \email{scao14@jhu.edu}
 \affiliation{%
   \institution{Johns Hopkins University}
   \streetaddress{3400 N. Charles St}
   \city{Baltimore}
   \state{Maryland}
   \country{USA}
   \postcode{21218}
 }

 \author{Laura Fleig}
 \email{lfleig1@jhu.edu}
 \affiliation{%
   \institution{Johns Hopkins University}
   \streetaddress{3400 N. Charles St}
   \city{Baltimore}
   \state{Maryland}
   \country{USA}
  \postcode{21218}
 }
 
 \author{Junan Jiang}
 \email{jjiang69@jh.edu}
 \affiliation{%
   \institution{Johns Hopkins University}
   \streetaddress{3400 N. Charles St}
   \city{Baltimore}
   \state{Maryland}
   \country{USA}
  \postcode{21218}
 }

  \author{Theodore Jensen}
 \email{theodore.jensen@nist.gov}
 \affiliation{%
   \institution{National Institute of Standards and Technology}
   \country{USA}
 }

 \author{Yee-Yin Choong}
 \email{yee-yin.choong@nist.gov}
 \affiliation{%
   \institution{National Institute of Standards and Technology}
   \country{USA}
 }

 \author{Kristen Greene}
 \email{kristen.greene@nist.gov}
 \affiliation{%
   \institution{National Institute of Standards and Technology}
   \country{USA}
 }

 \author{Chien-Ming Huang}
 \email{chienming.huang@jhu.edu}
 \authornote{Corresponding Author}
 \affiliation{%
   \institution{Johns Hopkins University}
   \streetaddress{3400 N. Charles St}
   \city{Baltimore}
   \state{Maryland}
   \country{USA}
   \postcode{21218}
 }

\begin{abstract}
Ambient artificial intelligence (AI) scribes have moved rapidly from experimental pilots to enterprise-wide deployment across health systems, promising reduced documentation burden and improved clinician experience. While research evaluating these tools has expanded quickly, there are no agreed-upon protocols or standardized metrics for assessing their performance, making it difficult to compare findings or cumulatively synthesize evidence across studies. Across 54 recent evaluation studies, we identify ten evaluation dimensions and 40 distinct metrics, operationalized in 96 different ways---35 of them (36\%) used by only a single study. Assessment is dominated by documentation efficiency and time, measured in 44 of 54 studies. The other dimensions appear at broadly similar rates (16–24 studies each), and safety and clinical outcome least consistently of all. Ripple effects on the broader care system, such as impact on patients' care outcomes, redistribution of work, changes in billing and coding intensity, and downstream use of notes by other clinicians are rarely assessed. Moreover, the metrics used rely heavily on self-report and are fragmented and inconsistently operationalized, which limits comparability and hinders synthesis across studies. As an effort to support more rigorous and cumulative evaluation, in this paper we review the study designs, data sources, and metrics commonly used to evaluate ambient AI scribes in healthcare settings, discuss the limitations of current evaluation practice, and argue for a more holistic approach grounded in a standardized, multi-dimensional set of metrics. 
\end{abstract}


%\ccsdesc[500]{Computing methodologies~Artificial intelligence}
%\ccsdesc[500]{Human-centered computing~Human computer interaction (HCI)}

% \keywords{AI benefit evaluation,  human-AI teaming, auditing,  organizational adoption, workflow integration, evaluation framework}


\maketitle

\section{Introduction}\label{sec:intro}
%\cmh{I'd remove the current first paragraph and get to healthcare right away. I'm also less sure about citing consulting firms' reports in peer-reviewed publications. } 
%Artificial intelligence (AI) is increasingly used to support decision-making, service delivery, and operational processes across a range of sectors. 
%Projections estimate that generative AI could raise global GDP by approximately 7\% (on the order of \$7~trillion) over a ten-year period and increase annual productivity growth by roughly 1.5 percentage points \cite{sachs2023generative}. Organization-level returns reported to date appear more limited: survey evidence indicates that relatively few organizations report a measurable enterprise-wide impact from AI, although most intend to increase their AI investment in the coming years \cite{singla2025state,mayer2025superagency}. The distance between projected value and demonstrated impact motivates the development of rigorous and practical methods for evaluating whether AI systems deliver their intended benefits. %\cite{ai2023artificial}

Clinical documentation is a substantial component of clinician workload. The widespread adoption of electronic health records (EHRs), driven by regulation such as the 2009 HITECH Act in the United States \cite{act2009health, halamka2017hitech}, has only increased the length and redundancy of progress notes \cite{rule2021length}. EHR-related documentation burden has been associated with clinician stress and burnout \cite{gardner2019physician,downing2018physician,wu2024evaluating}. Time-and-motion studies estimate that physicians spend approximately two hours on the electronic health record (EHR) and desk work for each hour of direct patient care, along with one to two hours of after-hours documentation each night \cite{sinsky2016allocation}. 
Reducing this burden---time spent in the EHR, after-hours (``pajama'') documentation, and the cognitive load of charting---has become a primary rationale for the introduction of ambient AI scribes.

Ambient AI scribes are systems that capture the clinician--patient encounter and generate a draft clinical note using automatic speech recognition and large language models. Over a period of a few years, these tools have progressed from small pilot studies to enterprise-wide deployment across several health systems \cite{holmgren2026ambient,wright2026enterprise,tierney2025ambient,rotenstein2026changes}. In contrast to earlier documentation aids, they require little change to the encounter itself: upon obtaining consent, the clinician speaks with the patient, and a structured note is produced for review, editing, and sign-off. The commercial landscape has expanded accordingly, with on the order of one hundred ambient-scribe products now available and substantial investment across the segment \cite{elion2025scribes,peterson2025adoption}. Procurement has been driven largely by the goal of mitigating clinician burnout, with gains in quality of care, patient experience, and financial performance cited as secondary aims, and has been facilitated by the tools' perceived low risk to patients, relatively straightforward integration with existing EHRs, and minimal workflow disruption \cite{peterson2025adoption}. Across these deployments, adoption has proceeded more quickly than the accumulation of evidence on the tools' effects.
 
Whether ambient scribes achieve these aims, and at what cost to note quality, patient experience, and safety, remains an empirical question that depends on how these tools are evaluated. Evaluating AI systems is methodologically challenging, and prior work has emphasized that constructs such as effectiveness, trust, and the quality of human-AI collaboration require deliberate measurement \cite{lai2023towards,vereschak2021evaluate,reverberi2022experimental}.
However, evaluation of ambient scribes have so far lacked shared standards, with individual studies select their own outcomes, instruments, and analytic designs. Even studies addressing the same construct, for example, burnout, cognitive load, or documentation time, frequently use different instruments and operational definitions.
 
As a result, the literature is substantial in volume but heterogeneous and difficult to synthesize. A time saving reported by one study cannot be readily placed alongside a saving defined over a different interval, and a change in burnout measured with a single-item scale cannot be combined with one measured using a multi-dimensional inventory. This limits comparison and cumulative synthesis at precisely the point where procurement, policy, and safety decisions about the technology are being made \cite{peterson2025adoption}.
In this review, we examine how ambient AI scribes are currently evaluated across study designs, data sources, and metrics, characterizing where evaluation is concentrated and where it is limited.
\am{Our question is not whether ambient scribes are effective, but whether current evaluation practice, in the constructs it measures and the methods it uses, can answer that question at all.}

This review makes four contributions. First, we provide a structured map of what is measured and how, organized by evaluation dimension and by instrument. Second, we quantify the concentration of measurement across dimensions, showing which are heavily evaluated and which are neglected. Third, we characterize the measurement quality problems that limit synthesis, including heavy reliance on self-report and inconsistent operationalization of even ostensibly objective EHR metrics. Fourth, we offer recommendations toward standardized, holistic, and multi-dimensional evaluation.

\section{Background}
Several reviews have synthesized the emerging evidence on ambient AI scribes in healthcare from the perspectives of effectiveness, implementation, and usability. Early systematic reviews primarily examined whether ambient AI scribes reduce documentation burden and improve clinical workflows, consistently reporting improvements in documentation efficiency and clinician engagement while noting limited and heterogeneous evidence regarding burnout reduction, patient outcomes, and organizational impact \cite{sasseville2025impact}. More recent reviews expanded the focus to implementation and usability, identifying reductions in cognitive workload and after-hours documentation alongside persistent challenges such as documentation errors, editing burden, limited customization, and incomplete EHR integration \cite{atiku2026usability}. Kanaparthy et al. further synthesized real-world evidence using the QUEST and SEIPS 3.0 frameworks, concluding that existing evaluations remain concentrated on workflow efficiency and clinician experience, whereas comparatively little evidence exists regarding safety, patient outcomes, productivity, and organizational performance \cite{kanaparthy2025real}. Overall, prior reviews have established what benefits and challenges of ambient AI scribes have been reported, but not how these findings have been generated.

More broadly, the evaluation of AI-enabled systems has increasingly emphasized multi-dimensional assessment beyond technical performance alone. Research on human-AI collaboration argues that evaluation should capture not only AI performance but also human factors, collaborative outcomes, workflow, and organizational context \cite{lai2023towards}. 
Several frameworks and reporting guidelines address how AI in healthcare should be evaluated. For instance, SEIPS \cite{carayon2006work} situates a technology within its wider work system and directs attention beyond patient outcomes to the clinicians and organizations affected and specifies possible relationships between patient outcomes and employee/organizational outcomes. Whereas, DECIDE-AI \cite{vasey2022reporting} specifies what the early, real-world evaluation of an AI intervention should report; it emphasizes clinical utility, safety, and human factors. The NIST AI Risk Management Framework (\textit{AI RMF 1.0}) \cite{nist_ai_rmf} offers broad, use-case-agnostic guidance for managing AI risks and promoting trustworthy systems across the AI lifecycle. These frameworks provide valuable high-level guidance on what should be considered when evaluating AI systems at scale, namely technical, human, implementation, safety, and organizational dimensions. Yet little is known about what the evaluation landscape actually looks like for AI now widely deployed in healthcare settings, such as ambient AI scribes; which evaluation dimensions have received the greatest attention, how similar constructs are operationalized across studies, whether commonly reported metrics are directly comparable, and where important evidence gaps remain. Without a systematic understanding of current evaluation practice, it is difficult to synthesize evidence across studies, support procurement and implementation decisions, or establish standardized evaluation practices for future deployments.

This review shifts the focus from what ambient AI scribes achieve to how those outcomes are evaluated. Rather than synthesizing further evidence on documentation efficiency, clinician workload, usability, or implementation, we examine the evaluation itself---the study designs, dimensions, metrics, data sources, instruments, metrics, and operational definitions behind the current evidence base. In doing so, we surface recurring patterns, inconsistencies, and underexamined dimensions, and lay the groundwork for more standardized, comprehensive, and multidimensional evaluation of ambient AI scribes.

\begingroup
\small
\setlength{\tabcolsep}{4pt}
\renewcommand{\arraystretch}{1.25}
\newcolumntype{L}[1]{>{\raggedright\arraybackslash}p{#1}}
\begin{longtable}{@{}L{4.0cm} L{12.0cm}@{}}
\caption{PICO(S) Elements \cite{page2021prisma}. }\label{tab:picos}\\
\toprule
\textbf{PICO(S) Elements} & \textbf{Elements in this Review} \\
\midrule
\endfirsthead
\multicolumn{2}{@{}l}{\small\emph{Table~\ref{tab:picos} continued}}\\
\toprule
\textbf{PICO(S) Elements} & \textbf{Elements in this Review} \\
\midrule
\endhead
\midrule\multicolumn{2}{r@{}}{\small\emph{continued on next page}}\\ \endfoot
\bottomrule\endlastfoot

Participants &
All healthcare providers engaged in clinical documentation \\
\addlinespace[1pt]

Intervention &
AI tools designed to streamline clinical documentation, including:
\begin{itemize}[nosep, leftmargin=*, itemsep=1pt]
  \item Real-time transcription and notetaking
  \item Automated data entry into EHRs
  \item Natural Language Processing (NLP) for clinical summarization, contextual understanding, and structured documentation
  \item Tools transforming spoken interactions into organized clinical notes
\end{itemize} \\
\addlinespace[1pt]

Comparison &
Comparison not required. \\
\addlinespace[1pt]

Outcomes &
All outcomes \\
\addlinespace[1pt]

Setting &
All clinical settings \\
\addlinespace[1pt]

Specifications for the studies &
\begin{itemize}[nosep, leftmargin=*, itemsep=1pt]
  \item Eligible studies: Included evaluation of a deployed ambient AI scribe in a clinical setting.
  \item Studies, pre-post studies, observational, qualitative-only studies, and mixed-method studies.
  \item Excluded studies: Not AI scribe, lack of real-world deployment, simulated patients, simulated setting, veterinary setting.
\end{itemize} \\
\addlinespace[1pt]

Language and publication year &
English-only; no restrictions. \\

\end{longtable}
\endgroup

\section{Methodology}\label{sec:methods}
% topic sentences
We conducted a systematic review reported in accordance with the PRISMA~2020 guidelines \cite{page2021prisma}. 

\subsection{Eligibility Criteria and Search Strategy}
Studies were eligible if they empirically evaluated the deployment of an AI ambient scribe in actual clinical practice involving real patients and reported at least one outcome.  We adapted our search strategy from a previous systematic review examining the effects of AI scribes on streamlining clinical documentation \cite{sasseville2025impact}. We excluded papers that are not written in English, evaluated human or virtual scribes rather than AI ambient scribes, did not involve a real-world clinical deployment, were conducted in simulated settings, or were performed in veterinary settings. Our full eligibility criteria follow the PICO(S) framework (Table~\ref{tab:picos}).

We searched Medline (OVID), Embase (Elsevier), CINAHL, Web of Science, arXiv, and Google Scholar for studies published between November 1, 2024 and July 31, 2026. The search combined the term ``AI scribe'' and a list of equivalent synonyms such as ``ambient scribe'', ``clinical scribe'', and ``medical scribe'' (full search strategy in Appendix~\ref{app:search}, Table \ref{tab:search-strategy}). We also included eight studies identified in the previous review \cite{sasseville2025impact}, which had used a broader search strategy to identify papers published before November 1, 2024. Together, our search was intended to capture all eligible studies published before July 31, 2026.

\subsection{Data Extraction}
We extracted study characteristics, designs, data sources, and every reported metric using a structured form. One reviewer completed the initial data extraction, and a second reviewer checked the extracted data for accuracy and completeness, with any discrepancies resolved by consensus. We then coded each formally measured metric into one of ten evaluation dimensions and, within each dimension, to the specific instrument or operationalization used, distinguishing validated instruments from self-established (bespoke) measures. Because outcomes and their operationalizations were highly heterogeneous, we synthesized the evidence narratively and through frequency counts rather than meta-analysis.
 
% --- PRISMA placeholder (replace with real diagram) ---
\begin{figure}[t]\centering
\includegraphics[width=\textwidth]{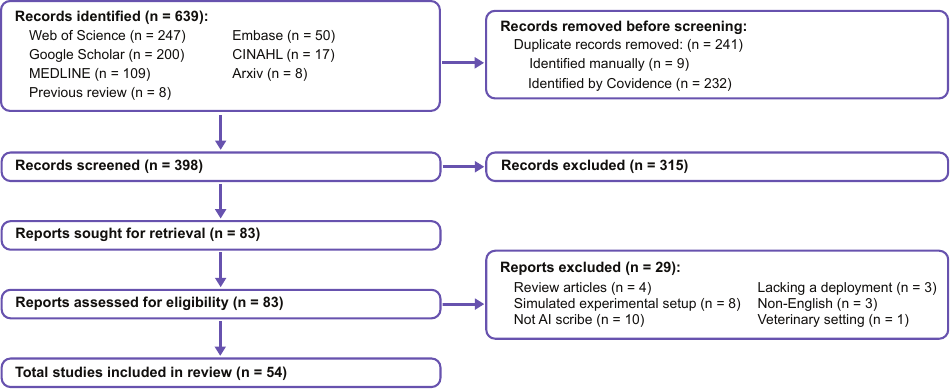}
\caption{PRISMA 2020 study selection flow.
}\label{fig:prisma}
\end{figure}
 
% ============================================================
\section{Results: Study Designs, Settings, and Data Sources}\label{sec:methods-settings}
After removing 241 duplicates (232 by Covidence and 9 manually), 398 titles and abstracts were screened independently. Following exclusion of 315 irrelevant records, 83 references remained for full-text review. Two reviewers independently screened records in Covidence and resolved disagreements by consensus, excluding 29 (4 review articles, 8 for simulated experimental setup, 10 for not evaluating an AI scribe (\eg evaluating human virtual scribe), 3 for lacking a deployment, 3 non-English, and 1 veterinary setting) and yielding 54 references suitable for inclusion in this review. A PRISMA 2020 flow diagram \cite{page2021prisma} of the study inclusion process is shown in Figure~\ref{fig:prisma}.

% topic sentences
Table~\ref{tab:summary} summarizes the design, setting, duration, methodology, and data sources of the 54 included studies.

\subsection{Study designs}
The evidence base is dominated by retrospective observational cohorts ($n=26$) and prospective/ quality improvement pilots ($n=25$), with only three randomized trials \cite{shin2025influence,lukac2025ambient,chowdhury2026comparing}, all of which ran 3--4 months. Assessing whether effects persist requires both a concurrent control and an extended follow-up window, yet no study combines the two beyond four months. %\cmh{why we talk about follow-up here? unclear to me}. 
Conversely, \textcolor{black}{21} of the \textcolor{black}{29} studies lasting longer than four months are retrospective observational cohort studies. In this literature, rigor and duration are inversely related, limiting inference about durable effects beyond an initial novelty period.
 
\subsection{Clinical settings and scale}
Evaluations cluster in ambulatory and primary-care settings, with comparatively little evidence from emergency, inpatient, or procedural specialties.
A handful of studies report enterprise-scale deployments involving hundreds of thousands to millions of encounters (\eg \cite{tierney2025ambient,holmgren2026ambient,waken2026longitudinal,sanmark2026evaluation,alcazar2026real}), but most remain single-clinic or single-department pilots. The concentration in ambulatory and primary care %\cmh{ambulatory and primary care or two different settings; is primary care ambulatory at all?} 
means that findings may not transfer to settings with different documentation demands, acuity, or workflows. 
 
\begingroup
\scriptsize
\setlength{\tabcolsep}{4pt}
\renewcommand{\arraystretch}{1.25}
\newcolumntype{L}[1]{>{\raggedright\arraybackslash}p{#1}}
\begin{longtable}{@{}L{7.5cm} c L{8cm}@{}}
\caption{Summary breakdown of the \textcolor{black}{54} reviewed studies by setting, duration, design, methodology, and data source. The middle three dimensions are mutually exclusive (each column sums to \textcolor{black}{54}). $^{\dagger}$Data sources are not mutually exclusive---many studies draw on several---so those counts exceed \textcolor{black}{54}. }\label{tab:summary}\\
\toprule
\textbf{Category} & \textbf{$n$} & \textbf{Studies} \\
\midrule
\endfirsthead
\multicolumn{3}{@{}l}{\scriptsize\emph{Table~\ref{tab:summary} continued}}\\
\toprule
\textbf{Category} & \textbf{$n$} & \textbf{Studies} \\
\midrule
\endhead
\midrule\multicolumn{3}{r@{}}{\scriptsize\emph{continued on next page}}\\ \endfoot
\bottomrule\endlastfoot
\rowcolor{gray!18}\multicolumn{3}{@{}l}{\textbf{Department / setting}}\\*
\quad Primary care / general practice /family medicine /internal medicine & \textcolor{black}{13} & \cite{toussi2026impact,ballantyne2025using,shah2025ambient,moura2026hybrid,north2025ambient,guo2025ambient,castro2026psychiatric,alpert2025evaluating,van2026ambient,waken2026longitudinal,memon2026performance,duggan2025clinician,husa2026ambient} \\
\quad Ambulatory \& multiple specialties (can include nurses, dentists etc.) & \textcolor{black}{25} & \cite{shah2025ambient,schneider2026impact,rotenstein2026changes,adler2026subjective,holmgren2026ambient,patel2026effect,wright2025effect,harvey2025ambient,wright2026enterprise,pearlman2025use,lukac2025ambient,olson2025use,chowdhury2026comparing, haberle2024impact, rui2025qualitative,guo2025ambient,evans2025impact, tierney2025ambient,tan2026impact,harvey2026ambient,wojda2026burden,sanmark2026evaluation,alcazar2026real,husa2026ambient,lawrence2026implementing} \\
\quad Pediatrics & \textcolor{black}{6} & \cite{shin2025influence,pelletier2025effect,rabbani2025ambient, north2025ambient,talman2026ambient,kapoor2026improving} \\
\quad Emergency medicine & \textcolor{black}{7} & \cite{wright2026enterprise,preiksaitis2026ambient,marquis2026ai,akhlaghi2026evaluation,dutta2026medical,alcazar2026deployment,preiksaitis2026ambientAI} \\
\quad Oncology / cancer care & \textcolor{black}{5} & \cite{toussi2026impact,nguyen2023implementing,baldwin2025ambient,baldwin2025early,talman2026ambient}\\
\quad Mental health & 1 & \cite{mccrudden2026ai} \\
\quad Dermatology & 1 & \cite{cao2024artificial} \\
\quad Orthopedic & \textcolor{black}{1} & \textcolor{black}{\cite{vallabhaneni2026reduction}} \\
\quad General surgery & \textcolor{black}{1} & \textcolor{black}{\cite{carmichael2026artificial}} \\
\quad Unspecified & \textcolor{black}{1} & \textcolor{black}{\cite{arko2026pre}} \\
\addlinespace[2pt]
\rowcolor{gray!18}\multicolumn{3}{@{}l}{\textbf{Study duration}}\\*
\quad $\leq$2 months & \textcolor{black}{7} & \cite{ballantyne2025using,nguyen2023implementing,wright2025effect,alpert2025evaluating,duggan2025clinician,akhlaghi2026evaluation,arko2026pre} \\
\quad 3--4 months & \textcolor{black}{18} & \cite{toussi2026impact,rui2025qualitative,adler2026subjective,shin2025influence,patel2026effect,shah2025ambient,moura2026hybrid,memon2026performance,harvey2025ambient,baldwin2025ambient,rabbani2025ambient,wright2026enterprise,pearlman2025use,lukac2025ambient,baldwin2025early,chowdhury2026comparing,harvey2026ambient,carmichael2026artificial} \\
\quad 5--9 months & \textcolor{black}{15} & \cite{haberle2024impact,schneider2026impact,guo2025ambient,pelletier2025effect,evans2025impact,preiksaitis2026ambient,van2026ambient,olson2025use,tan2026impact,marquis2026ai,talman2026ambient,wojda2026burden,kapoor2026improving,lawrence2026implementing,dutta2026medical} \\
\quad $\sim$1 year or more & \textcolor{black}{14} & \cite{rotenstein2026changes,holmgren2026ambient,mccrudden2026ai,north2025ambient,cao2024artificial,castro2026psychiatric,tierney2025ambient,waken2026longitudinal,sanmark2026evaluation,alcazar2026real,husa2026ambient,vallabhaneni2026reduction,alcazar2026deployment,preiksaitis2026ambientAI} \\
\addlinespace[2pt]
\rowcolor{gray!18}\multicolumn{3}{@{}l}{\textbf{Study type (design)}}\\*
\quad Prospective / quality-improvement pilot & \textcolor{black}{25} & \cite{toussi2026impact,nguyen2023implementing,schneider2026impact,shah2025ambient,guo2025ambient,pelletier2025effect,cao2024artificial,memon2026performance,wright2025effect,harvey2025ambient,baldwin2025ambient,rabbani2025ambient,alpert2025evaluating,baldwin2025early,olson2025use,duggan2025clinician,marquis2026ai,harvey2026ambient,arko2026pre,lawrence2026implementing,carmichael2026artificial, ballantyne2025using,rui2025qualitative,patel2026effect,wojda2026burden} \\
\quad Retrospective observational / cohort & \textcolor{black}{26} & \cite{van2026ambient,rotenstein2026changes,haberle2024impact,adler2026subjective,holmgren2026ambient,moura2026hybrid,north2025ambient,mccrudden2026ai,castro2026psychiatric,preiksaitis2026ambient,tierney2025ambient,pearlman2025use,evans2025impact,wright2026enterprise,waken2026longitudinal,tan2026impact,akhlaghi2026evaluation,talman2026ambient,sanmark2026evaluation,alcazar2026real,husa2026ambient,vallabhaneni2026reduction,kapoor2026improving,dutta2026medical,alcazar2026deployment,preiksaitis2026ambientAI} \\
\quad Randomized trial (incl.\ crossover) & 3 & \cite{shin2025influence,lukac2025ambient,chowdhury2026comparing} \\
%\quad Cross-sectional survey / exploratory & \textcolor{black}{4} & \cite{ballantyne2025using,rui2025qualitative,patel2026effect,wojda2026burden} \\

\addlinespace[2pt]
\rowcolor{gray!18}\multicolumn{3}{@{}l}{\textbf{Methodology}}\\*
\quad Quantitative only & \textcolor{black}{35} & 
\cite{adler2026subjective,cao2024artificial,castro2026psychiatric,guo2025ambient,haberle2024impact,harvey2025ambient,holmgren2026ambient,lukac2025ambient,moura2026hybrid,north2025ambient,olson2025use,patel2026effect,pearlman2025use,preiksaitis2026ambient,rotenstein2026changes,schneider2026impact,shah2025ambient,shin2025influence,tierney2025ambient,toussi2026impact,waken2026longitudinal,wright2025effect,akhlaghi2026evaluation,harvey2026ambient,wojda2026burden,arko2026pre,alcazar2026real,husa2026ambient,vallabhaneni2026reduction,kapoor2026improving,carmichael2026artificial,dutta2026medical,alcazar2026deployment,preiksaitis2026ambientAI,sanmark2026evaluation}
\\
\quad Mixed-methods (quant.\ + qual.) & \textcolor{black}{19} & 
\cite{alpert2025evaluating,baldwin2025ambient,baldwin2025early, ballantyne2025using,evans2025impact,van2026ambient, memon2026performance,nguyen2023implementing,pelletier2025effect,rabbani2025ambient,rui2025qualitative,tan2026impact,wright2026enterprise,chowdhury2026comparing,duggan2025clinician,marquis2026ai,mccrudden2026ai,talman2026ambient,lawrence2026implementing}
\\
\addlinespace[2pt]
\rowcolor{gray!18}\multicolumn{3}{@{}l}{\textbf{Data sources used $^{\dagger}$}}\\*
\quad EHR / audit-log metrics (Epic Signal, EHR records, productivity) & \textcolor{black}{39} & \cite{van2026ambient,evans2025impact,rotenstein2026changes,haberle2024impact,adler2026subjective,shin2025influence,holmgren2026ambient,mccrudden2026ai,moura2026hybrid,north2025ambient,guo2025ambient,pelletier2025effect,cao2024artificial,wright2025effect,harvey2025ambient,baldwin2025ambient,rabbani2025ambient,castro2026psychiatric,wright2026enterprise,preiksaitis2026ambient,alpert2025evaluating,tierney2025ambient,pearlman2025use,lukac2025ambient,waken2026longitudinal,baldwin2025early,chowdhury2026comparing,duggan2025clinician,talman2026ambient,harvey2026ambient,arko2026pre,alcazar2026real,husa2026ambient,vallabhaneni2026reduction,kapoor2026improving,lawrence2026implementing,dutta2026medical,alcazar2026deployment,preiksaitis2026ambientAI} \\
\quad Survey / questionnaire responses & 40 & \cite{ballantyne2025using,toussi2026impact,nguyen2023implementing,haberle2024impact,rui2025qualitative,adler2026subjective,shin2025influence,patel2026effect,schneider2026impact,mccrudden2026ai,shah2025ambient,pelletier2025effect,cao2024artificial,memon2026performance,wright2025effect,harvey2025ambient,evans2025impact,baldwin2025ambient,rabbani2025ambient,wright2026enterprise,alpert2025evaluating,tierney2025ambient,lukac2025ambient,baldwin2025early,olson2025use,tan2026impact,marquis2026ai,chowdhury2026comparing,duggan2025clinician,talman2026ambient,harvey2026ambient,wojda2026burden,sanmark2026evaluation,arko2026pre,alcazar2026real,kapoor2026improving,lawrence2026implementing,alcazar2026deployment,dutta2026medical,van2026ambient}
\\
\quad Interview transcripts & \textcolor{black}{8} & \cite{nguyen2023implementing,memon2026performance,evans2025impact,baldwin2025ambient,alpert2025evaluating,van2026ambient,baldwin2025early,lawrence2026implementing} \\
\quad Directly observed consultations & \textcolor{black}{3} & \cite{van2026ambient,tan2026impact, carmichael2026artificial} \\
\quad Clinical notes / documents (length, PDQI, ROUGE, text) & 20 & \cite{mccrudden2026ai,north2025ambient,guo2025ambient,cao2024artificial,memon2026performance,harvey2025ambient,castro2026psychiatric,preiksaitis2026ambient,alpert2025evaluating,van2026ambient,waken2026longitudinal,duggan2025clinician,akhlaghi2026evaluation,sanmark2026evaluation,alcazar2026real,alcazar2026deployment,harvey2026ambient,kapoor2026improving,dutta2026medical,preiksaitis2026ambientAI}
\\
\quad Billing / claims / RVU data & \textcolor{black}{10} & \cite{rotenstein2026changes,toussi2026impact,haberle2024impact,shin2025influence,holmgren2026ambient,moura2026hybrid,harvey2025ambient,waken2026longitudinal,husa2026ambient,dutta2026medical} \\
\addlinespace[2pt]
\end{longtable}
\endgroup
\subsection{Data sources}
\label{sec:datasources}
Studies drew evidence primarily from two sources: clinician self-report via surveys (40 studies) and EHR audit-log metrics (\textcolor{black}{39} studies). In contrast, interview transcripts (\textcolor{black}{8} studies), clinical-note or document analysis (20 studies), billing or claims data (\textcolor{black}{10} studies), and direct observation of consultations (\textcolor{black}{3} studies) were far less common. Notably, no study relied on qualitative methods alone, and \textcolor{black}{35} of \textcolor{black}{54} studies used quantitative methods only. 27 studies used both EHR audit logs and surveys; 12 used EHR only; \textcolor{black}{13} used survey only; and \textcolor{black}{two} used neither. The dominance of survey and audit-log data means that most of what is known about ambient scribes rests either on what clinicians report or on what the EHR happens to log, rather than on direct assessment of the notes the tools produce or the encounters they affect. Direct note audits and in-person observation are, admittedly, more resource-intensive and raise privacy considerations given the sensitivity of clinical encounters, which partly explains their rarity; they nonetheless remain the only sources that assess the note itself or the encounter directly.

% ============================================================
\section{Results: Evaluation Metrics}\label{sec:metrics}
\begin{figure}[b]\centering
\includegraphics[width=\textwidth]{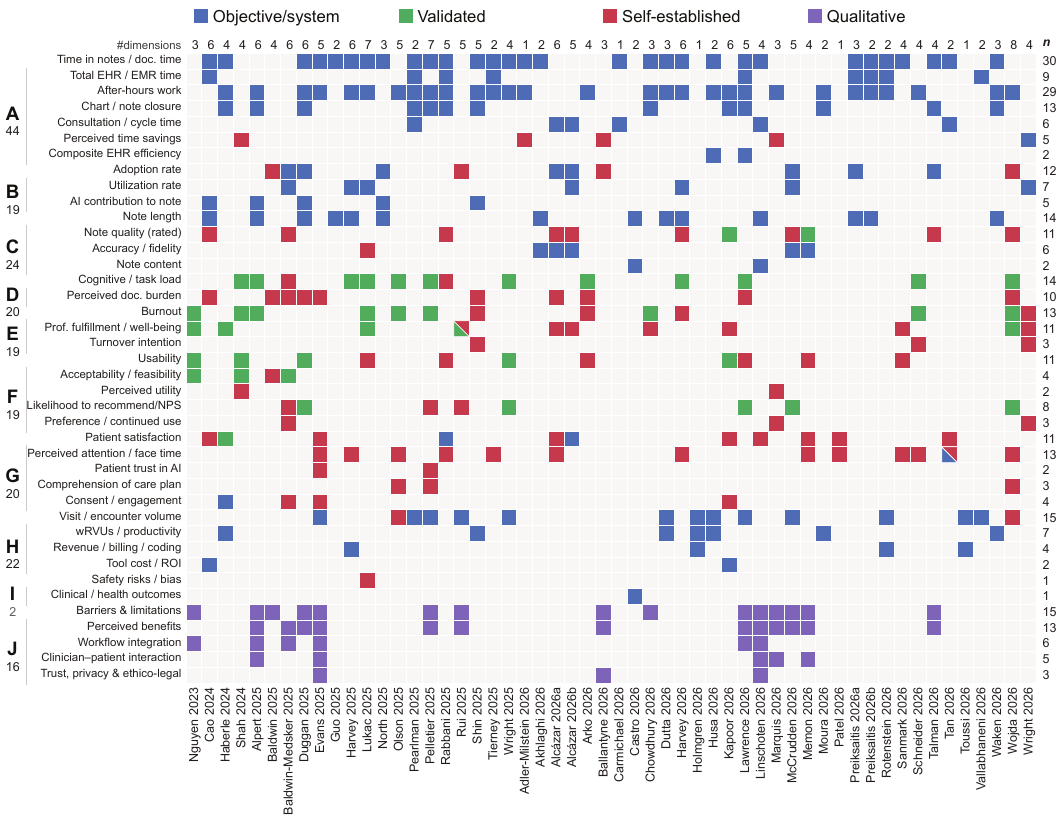}
\caption{ Metric-by-study measurement map. Rows are the metrics grouped into the ten evaluation dimensions (A: efficiency \& time; B: adoption \& utilization; C: note characteristics, quality \& accuracy; D: cognitive load \& burden; E: well-being, burnout \& turnover; F: usability \& acceptability; G: patient experience; H: productivity, billing \& financial; I: safety \& clinical outcomes; J: qualitative concepts); the count beneath each dimension letter ($n$) is the number of studies measuring that dimension, and the $n$ column at the right is the number measuring each metric. Columns are the \textcolor{black}{54} studies, \textcolor{black}{ordered by publication year (2023--2026)}; the \#dimensions row along the top gives how many of the ten dimensions each study covers (range 1--\textcolor{black}{8}). Cell colour denotes the type of instrument used (objective/system, validated, self-established, or qualitative). 
 The sparse, non-overlapping pattern shows that studies rarely share the same measurement footprint.}
\label{fig:map}
\end{figure}

\textcolor{black}{The 54 studies measured 40 distinct metrics, operationalized in 96 different ways, across ten evaluation dimensions, averaging nearly two distinct operationalizations per study.} Table~\ref{tab:metrics} maps these metrics to the ten dimensions \textcolor{black}{that we synthesized inductively from the extracted metrics} and to the specific instruments used within each. 
\am{Some of this diversity is structural, as different EHR platforms expose different fields and compute the same construct differently, so a metric such as ``time in notes'' is often defined by whatever a given system happens to log rather than by a shared standard.} 
Measurement is concentrated in a few dimensions, including documentation efficiency, note characteristics, adoption, and clinician cognitive load. Safety, and downstream effects such as patient comprehension, productivity, and billing are measured far less often (Figure~\ref{fig:map}).

 % Metrics table: metric -> measurement method -> studies (Author-year \cite).
% Named/validated instruments each get a row; self-established items folded per construct.
% Requires: booktabs, longtable, array, multirow, xcolor; natbib author-year.
% RED = added/changed in this revision (14 new studies; corpus 39 -> 53).
\begingroup
\scriptsize
\setlength{\tabcolsep}{4pt}
\renewcommand{\arraystretch}{1.32}
\newcolumntype{L}[1]{>{\raggedright\arraybackslash}p{#1}}
\begin{longtable}{@{}L{4cm} L{7.5cm} c L{4cm}@{}}
\caption{Metrics used to evaluate ambient AI scribes across \textcolor{black}{54} studies, grouped by evaluation theme and broken down by the specific instrument or operationalization. Each validated / named instrument is listed separately; heterogeneous self-established (bespoke) items for the same construct are grouped into a single row. ``$n$'' is the number of studies using that particular measurement; the metric total (unique studies) appears beside each metric name. }\label{tab:metrics}\\
\toprule
\textbf{Metric} & \textbf{Measurement / instrument} & \textbf{$n$} & \textbf{Studies} \\
\midrule
\endfirsthead
\multicolumn{4}{@{}l}{\scriptsize\emph{Table~\ref{tab:metrics} continued from previous page}}\\
\toprule
\textbf{Metric} & \textbf{Measurement / instrument} & \textbf{$n$} & \textbf{Studies} \\
\midrule
\endhead
\midrule\multicolumn{4}{r@{}}{\scriptsize\emph{continued on next page}}\\ \endfoot
\bottomrule\endlastfoot
\rowcolor{gray!18}
\multicolumn{4}{@{}l}{\textbf{A.~Efficiency \& time}\quad\textup{\scriptsize(\textcolor{black}{44} studies)}}\\*
\multirow[t]{3}{=}{\raggedright Time in notes / documentation time\newline\textit{\scriptsize(n=\textcolor{black}{30})}} & EHR / Epic Signal ``time in notes'' (per note, appointment, day, weekly, or monthly average) & 20 & \cite{shin2025influence,north2025ambient,guo2025ambient,cao2024artificial,wright2025effect,harvey2025ambient,rabbani2025ambient,preiksaitis2026ambient,tierney2025ambient,pearlman2025use,lukac2025ambient,waken2026longitudinal,chowdhury2026comparing,duggan2025clinician,talman2026ambient,harvey2026ambient,husa2026ambient,lawrence2026implementing,dutta2026medical,preiksaitis2026ambientAI} \\*
 & DocTime (EHR minutes per scheduled hour) & 2 & \cite{rotenstein2026changes,adler2026subjective} \\*
 & Average documentation time per patient / consultation (bespoke timing or self-report) & 8 & \cite{haberle2024impact,van2026ambient,evans2025impact,pelletier2025effect,akhlaghi2026evaluation,carmichael2026artificial,sanmark2026evaluation,tan2026impact}
 \\
\cmidrule(l{0pt}r{0pt}){1-4}
\multirow[t]{2}{=}{\raggedright Total EHR / EMR time\newline\textit{\scriptsize(n=\textcolor{black}{9})}} & Aggregate EHR minutes per day or appointment (Epic Signal; EHR-8 in Rotenstein) & \textcolor{black}{8} & \cite{rotenstein2026changes,cao2024artificial,rabbani2025ambient,preiksaitis2026ambient,pearlman2025use,tierney2025ambient,vallabhaneni2026reduction,preiksaitis2026ambientAI} \\*
 & Time in specific EHR task categories (in-basket, orders, clinical review, per appt) & \textcolor{black}{2} & \cite{tierney2025ambient,lawrence2026implementing} \\
\cmidrule(l{0pt}r{0pt}){1-4}
\multirow[t]{5}{=}{\raggedright After-hours work\newline\textit{\scriptsize(n=\textcolor{black}{29})}} & Pajama time (Epic: charting outside work hours on weekdays, weekends, holidays) & \textcolor{black}{9} & \cite{shin2025influence,harvey2025ambient,tierney2025ambient,waken2026longitudinal,chowdhury2026comparing,duggan2025clinician,wright2025effect,harvey2026ambient,lawrence2026implementing} \\*
 & Time Outside Scheduled Hours (TOSH) & 6 & \cite{rabbani2025ambient,alpert2025evaluating,lukac2025ambient,waken2026longitudinal,chowdhury2026comparing,wright2025effect} \\*
 & Work Outside Work (WOW) & 4 & \cite{rotenstein2026changes,adler2026subjective,moura2026hybrid,tierney2025ambient} \\*
 & Time in EHR on unscheduled days & 4 & \cite{alpert2025evaluating,lukac2025ambient,tierney2025ambient,wright2025effect} \\*
 & After-shift / after-hours documentation (defined window or self-report) & \textcolor{black}{15} & \cite{marquis2026ai,haberle2024impact,preiksaitis2026ambient,evans2025impact,pearlman2025use,olson2025use, schneider2026impact, pelletier2025effect,harvey2026ambient,wojda2026burden,arko2026pre,husa2026ambient,kapoor2026improving,dutta2026medical,preiksaitis2026ambientAI} \\
\cmidrule(l{0pt}r{0pt}){1-4}
\multirow[t]{3}{=}{\raggedright Chart / note closure timeliness\newline\textit{\scriptsize(n=\textcolor{black}{13})}} & Same-day chart / visit closure rate & \textcolor{black}{7} & \cite{shin2025influence,pelletier2025effect,rabbani2025ambient,alpert2025evaluating,duggan2025clinician,talman2026ambient,lawrence2026implementing} \\*
 & Notes signed / closed within 24h, 48h, or 3 business days (documentation delay) & \textcolor{black}{5} & \cite{haberle2024impact,moura2026hybrid,waken2026longitudinal,chowdhury2026comparing,lawrence2026implementing} \\*
 & Time to close encounters & \textcolor{black}{2} & \cite{pearlman2025use,kapoor2026improving} \\
\cmidrule(l{0pt}r{0pt}){1-4}
\multirow[t]{2}{=}{\raggedright Consultation / cycle time\newline\textit{\scriptsize(n=\textcolor{black}{6})}} & Total consultation / appointment duration & \textcolor{black}{6} & \cite{van2026ambient,pearlman2025use,tan2026impact,alcazar2026real,carmichael2026artificial,alcazar2026deployment} \\*
 & Total patient cycle time & 1 & \cite{tan2026impact} \\
\cmidrule(l{0pt}r{0pt}){1-4}
\multirow[t]{2}{=}{\raggedright Perceived / estimated time savings\newline\textit{\scriptsize(n=5)}} & Self-established item: time saved (survey) & 4 & \cite{ballantyne2025using,marquis2026ai,shah2025ambient,adler2026subjective} \\*
 & Estimated time saving per encounter & 1 & \cite{wright2026enterprise} \\
\cmidrule(l{0pt}r{0pt}){1-4}
\multirow[t]{1}{=}{\raggedright \textcolor{black}{Composite EHR efficiency score}\newline\textit{\scriptsize\textcolor{black}{(n=2)}}} & \textcolor{black}{Vendor/Epic composite (Clinician Efficiency Profile [CEP], Provider Efficiency Profile [PEP], proficiency score)} & \textcolor{black}{2} & \cite{husa2026ambient,lawrence2026implementing} \\
\cmidrule(l{0pt}r{0pt}){1-4}
\rowcolor{gray!18}
\multicolumn{4}{@{}l}{\textbf{B.~Adoption \& utilization}\quad\textup{\scriptsize(\textcolor{black}{19} studies)}}\\*
\multirow[t]{2}{=}{\raggedright Adoption rate (clinician-level)\newline\textit{\scriptsize(n=\textcolor{black}{12})}} & Percentage of eligible clinicians using the scribe (often weekly; incl.\ adoption concentration) & \textcolor{black}{8} & \cite{mccrudden2026ai,north2025ambient,preiksaitis2026ambient,baldwin2025ambient,duggan2025clinician,talman2026ambient,alcazar2026real,alcazar2026deployment} \\*
 & Self-established item: self-reported usage frequency / proportion bands & \textcolor{black}{4} & \cite{rui2025qualitative,ballantyne2025using,baldwin2025early,wojda2026burden} \\
\cmidrule(l{0pt}r{0pt}){1-4}
\multirow[t]{1}{=}{\raggedright Utilization rate (note / visit-level) \textit{\scriptsize(n=\textcolor{black}{7})}} & Percentage of notes or visits generated with the scribe (AI notes / total) & \textcolor{black}{7} & \cite{mccrudden2026ai,harvey2025ambient,wright2026enterprise,lukac2025ambient,baldwin2025ambient,harvey2026ambient,alcazar2026deployment} \\
\cmidrule(l{0pt}r{0pt}){1-4}
\multirow[t]{3}{=}{\raggedright AI contribution to note\newline\textit{\scriptsize(n=5)}} & Percent of note characters generated by AI & 2 & \cite{north2025ambient,cao2024artificial} \\*
 & Percent of characters manually typed / entered by clinician & 2 & \cite{shin2025influence,duggan2025clinician} \\*
 & Characters typed (count) & 1 & \cite{alpert2025evaluating} \\
\cmidrule(l{0pt}r{0pt}){1-4}
\rowcolor{gray!18}
\multicolumn{4}{@{}l}{\textbf{C.~Note characteristics, quality \& accuracy}\quad\textup{\scriptsize(\textcolor{black}{24} studies)}}\\*
\multirow[t]{3}{=}{\raggedright Note length\newline\textit{\scriptsize(n=\textcolor{black}{14})}} & Length in characters (summed / averaged over notes or physician-weeks) & \textcolor{black}{11} & \cite{north2025ambient,guo2025ambient,harvey2025ambient,castro2026psychiatric,preiksaitis2026ambient,alpert2025evaluating,waken2026longitudinal,duggan2025clinician,harvey2026ambient,dutta2026medical,preiksaitis2026ambientAI} \\*
 & Length in words & \textcolor{black}{2} & \cite{cao2024artificial,akhlaghi2026evaluation} \\*
 & Length by SOAP section & 1 & \cite{van2026ambient} \\
\cmidrule(l{0pt}r{0pt}){1-4}
\multirow[t]{3}{=}{\raggedright Note quality (rated)\newline\textit{\scriptsize(n=\textcolor{black}{11})}} & PDQI-9 (validated Physician Documentation Quality Instrument; 5-pt Likert, 8--9 domains, max 40) & \textcolor{black}{2} & \cite{memon2026performance,kapoor2026improving} \\*
 & Clinician thumbs-up / down rating with negative-reason coding & 1 & \cite{mccrudden2026ai} \\*
 & Self-established Likert note-quality / accuracy items & \textcolor{black}{8} & \cite{rabbani2025ambient,baldwin2025ambient,cao2024artificial,talman2026ambient,harvey2026ambient,wojda2026burden,alcazar2026real,alcazar2026deployment} \\
\cmidrule(l{0pt}r{0pt}){1-4}
\multirow[t]{5}{=}{\raggedright Accuracy / fidelity\newline\textit{\scriptsize(n=\textcolor{black}{6})}} & ROUGE-1 F1 vs.\ final medical record & 1 & \cite{memon2026performance} \\*
 & Dice--S{\o}rensen text similarity + generation error rate per week & 1 & \cite{mccrudden2026ai} \\*
 & Self-established item: clinician-reported inaccuracies and bias & 1 & \cite{lukac2025ambient} \\*
 & \textcolor{black}{Cosine similarity / semantic agreement between AI output and physician-validated note} & \textcolor{black}{2} & \textcolor{black}{\cite{alcazar2026real,alcazar2026deployment}} \\*
 & \textcolor{black}{Clinician edit distance (words / characters added or deleted vs.\ AI draft)} & \textcolor{black}{1} & \textcolor{black}{\cite{akhlaghi2026evaluation}} \\
\cmidrule(l{0pt}r{0pt}){1-4}
\multirow[t]{2}{=}{\raggedright Note content / completeness\newline\textit{\scriptsize(n=2)}} & Count of required note variables (signs, context, symptoms, diagnosis, plan) & 1 & \cite{van2026ambient} \\*
 & RDoC scores estimated from note text & 1 & \cite{castro2026psychiatric} \\
\cmidrule(l{0pt}r{0pt}){1-4}
\rowcolor{gray!18}
\multicolumn{4}{@{}l}{\textbf{D.~Cognitive load \& documentation burden}\quad\textup{\scriptsize(\textcolor{black}{20} studies)}}\\*
\multirow[t]{4}{=}{\raggedright Cognitive / task load\newline\textit{\scriptsize(n=\textcolor{black}{14})}} & NASA Task Load Index (NASA-TLX) (full or subscales: mental / physical demand, pace, effort, hurried, workload) & \textcolor{black}{10} & \cite{pelletier2025effect,schneider2026impact,harvey2025ambient,alpert2025evaluating,shah2025ambient,wright2025effect,harvey2026ambient,wojda2026burden,arko2026pre,lawrence2026implementing} \\*
 & Modified NASA-TLX & 1 & \cite{olson2025use} \\*
 & 4-item Provider Task Load (PTL; 0--400) & 1 & \cite{lukac2025ambient} \\*
 & Self-established mental-demand / cognitive-burden Likert item & 2 & \cite{baldwin2025ambient,rabbani2025ambient} \\
\cmidrule(l{0pt}r{0pt}){1-4}
\multirow[t]{1}{=}{\raggedright Perceived documentation burden/ stress \newline\textit{\scriptsize(n=\textcolor{black}{10})}} 
 & Self-established Likert burden items (effort, difficulty closing charts, burden of documentation) & \textcolor{black}{10} & \cite{evans2025impact,shin2025influence,duggan2025clinician,baldwin2025ambient,baldwin2025early,cao2024artificial,wojda2026burden,arko2026pre,alcazar2026real,lawrence2026implementing} \\
\cmidrule(l{0pt}r{0pt}){1-4}
\rowcolor{gray!18}
\multicolumn{4}{@{}l}{\textbf{E.~Clinician well-being, burnout \& turnover}\quad\textup{\scriptsize(\textcolor{black}{19} studies)}}\\*
\multirow[t]{4}{=}{\raggedright Burnout\newline\textit{\scriptsize(n=\textcolor{black}{13})}} & Mini-Z / Mini-Z 2.0 single-item; also validated against Maslach Burnout Inventory (MBI) & \textcolor{black}{7} & \cite{pelletier2025effect,nguyen2023implementing,alpert2025evaluating,lukac2025ambient,olson2025use, schneider2026impact,wojda2026burden} \\*
 & Copenhagen Burnout Inventory (CBI) subscales & 1 & \cite{chowdhury2026comparing} \\*
 &  Stanford Professional Fulfillment Index subscale work exhaustion (PFI-WE)  & 1 & \cite{shah2025ambient} \\*
 & Self-established burnout / fatigue item & \textcolor{black}{4} & \cite{shin2025influence,wright2026enterprise,harvey2026ambient,arko2026pre} \\
\cmidrule(l{0pt}r{0pt}){1-4}
\multirow[t]{4}{=}{\raggedright Professional fulfillment / well-being\newline\textit{\scriptsize(n=\textcolor{black}{11})}} & Professional Fulfillment Index (PFI / PFI--Work Exhaustion) & \textcolor{black}{3} & \cite{rui2025qualitative,lukac2025ambient,wojda2026burden} \\*
 & Press Ganey engagement / resilience / work-life balance & 1 & \cite{haberle2024impact} \\*
 & PROMIS SF-4a sleep disturbance & 1 & \cite{nguyen2023implementing} \\*
 & Self-established well-being / EHR-experience / satisfaction items & \textcolor{black}{7} & \cite{rui2025qualitative,wright2026enterprise,chowdhury2026comparing,sanmark2026evaluation,alcazar2026real,kapoor2026improving,alcazar2026deployment} \\
\cmidrule(l{0pt}r{0pt}){1-4}
\multirow[t]{1}{=}{\raggedright Turnover intention\newline\textit{\scriptsize(n=3)}} & Self-established item: likelihood of leaving the organization / clinical practice & 3 & \cite{shin2025influence,schneider2026impact,wright2026enterprise} \\
\cmidrule(l{0pt}r{0pt}){1-4}
\rowcolor{gray!18}
\multicolumn{4}{@{}l}{\textbf{F.~Usability \& acceptability}\quad\textup{\scriptsize(\textcolor{black}{19} studies)}}\\*
\multirow[t]{3}{=}{\raggedright Usability\newline\textit{\scriptsize(n=\textcolor{black}{11})}} & System Usability Scale (SUS; 10-item, validated) & \textcolor{black}{4} & \cite{shah2025ambient,duggan2025clinician, wright2025effect,kapoor2026improving} \\*
 & Intervention Usability Scale (0--100, validated) & 1 & \cite{nguyen2023implementing} \\*
 & Self-established usability / experience survey items & \textcolor{black}{6} & \cite{rabbani2025ambient,lukac2025ambient,memon2026performance,sanmark2026evaluation,arko2026pre,lawrence2026implementing} \\
\cmidrule(l{0pt}r{0pt}){1-4}
\multirow[t]{4}{=}{\raggedright Acceptability / feasibility / appropriateness\newline\textit{\scriptsize(n=4)}} & Weiner implementation-outcome measures (Acceptability, Feasibility, Appropriateness; composite 4--20, validated) & 1 & \cite{nguyen2023implementing} \\*
 & Proctor implementation-outcomes framework (adoption, feasibility, acceptability, appropriateness) & 1 & \cite{baldwin2025ambient} \\*
 & adapted and shortened version of the Unified Theory of Acceptance and Use of Technology (UTAUT) & 1 & \cite{shah2025ambient} \\*
 & Self-established acceptability item (unspecified instrument) & 1 & \cite{baldwin2025early} \\
\cmidrule(l{0pt}r{0pt}){1-4}
\multirow[t]{1}{=}{\raggedright Perceived utility / helpfulness\newline\textit{\scriptsize(n=2)}} & Self-established utility / helpfulness items (overall; for documentation; for physical-exam documentation) & 2 & \cite{marquis2026ai,shah2025ambient} \\
\cmidrule(l{0pt}r{0pt}){1-4}
\multirow[t]{2}{=}{\raggedright Likelihood to recommend / NPS\newline\textit{\scriptsize(n=\textcolor{black}{8})}} & Net Promoter Score & \textcolor{black}{5} & \cite{mccrudden2026ai,duggan2025clinician,wright2025effect,wojda2026burden,lawrence2026implementing} \\*
 & Self-established single-item likelihood to recommend to a colleague & 3 & \cite{rui2025qualitative,pelletier2025effect,baldwin2025ambient} \\
\cmidrule(l{0pt}r{0pt}){1-4}
\multirow[t]{1}{=}{\raggedright Preference / continued use / perceived value %\cmh{perceived value is different from perceived utility?} 
\textit{\scriptsize(n=3)}} & Self-established items: preference, intention to continue, trust in tool, feeling if tool unavailable & 3 & \cite{marquis2026ai,baldwin2025ambient,wright2026enterprise} \\
\cmidrule(l{0pt}r{0pt}){1-4}
\rowcolor{gray!18}
\multicolumn{4}{@{}l}{\textbf{G.~Patient experience}\quad\textup{\scriptsize(\textcolor{black}{20} studies)}}\\*
\multirow[t]{3}{=}{\raggedright Patient satisfaction / experience (overall)\newline\textit{\scriptsize(n=\textcolor{black}{11})}} & Press Ganey / Likelihood-to-Recommend (LTR) & 1 & \cite{haberle2024impact} \\*
 & Patient-experience survey incl.\ Net Promoter Score (NPS) & \textcolor{black}{2} & \cite{rabbani2025ambient,alcazar2026deployment} \\*
 & Self-established patient-satisfaction / experience survey items & \textcolor{black}{8} & \cite{patel2026effect,evans2025impact,tan2026impact,memon2026performance,van2026ambient,cao2024artificial,kapoor2026improving,alcazar2026real} \\
\cmidrule(l{0pt}r{0pt}){1-4}
\multirow[t]{4}{=}{\raggedright Perceived attention / face time\newline\textit{\scriptsize(n=\textcolor{black}{13})}} & Self-established item: ``doctor listens carefully / felt closely listened to''  & \textcolor{black}{5} & \cite{tan2026impact,harvey2025ambient,evans2025impact,rabbani2025ambient,harvey2026ambient} \\*
 & Self-established item: undivided / focused attention to patient (clinician-reported) & \textcolor{black}{5} & \cite{tierney2025ambient,olson2025use,wojda2026burden,sanmark2026evaluation,alcazar2026real} \\*
 & Self-established item: face time / time doctor spent / screen time & 3 & \cite{schneider2026impact,memon2026performance,patel2026effect} \\*
 & Eye contact (observed) & 1 & \cite{tan2026impact} \\
\cmidrule(l{0pt}r{0pt}){1-4}
\multirow[t]{1}{=}{\raggedright Patient trust in AI / clinician\textit{\scriptsize(n=2)}} & Self-established item: patient-reported trust in AI and in the clinician & 2 & \cite{evans2025impact,pelletier2025effect} \\
\cmidrule(l{0pt}r{0pt}){1-4}
\multirow[t]{1}{=}{\raggedright Comprehension of care plan \textit{\scriptsize(n=\textcolor{black}{3})}} & Self-established item: understandability of care plan from the note & \textcolor{black}{3} & \cite{pelletier2025effect,olson2025use,wojda2026burden} \\
\cmidrule(l{0pt}r{0pt}){1-4}
\multirow[t]{3}{=}{\raggedright Consent / engagement / return intent\newline\textit{\scriptsize(n=\textcolor{black}{4})}} & Patient opt-out rate & 1 & \cite{haberle2024impact} \\*
 & Self-established item: impact on future-appointment decisions & 1 & \cite{evans2025impact} \\*
 & Self-established item: clinician-rated patient engagement & \textcolor{black}{2} & \cite{baldwin2025ambient,kapoor2026improving} \\
\cmidrule(l{0pt}r{0pt}){1-4}
\rowcolor{gray!18}
\multicolumn{4}{@{}l}{\textbf{H.~Productivity, billing \& financial}\quad\textup{\scriptsize(\textcolor{black}{22} studies)}}\\*
\multirow[t]{5}{=}{\raggedright Visit / encounter volume\newline\textit{\scriptsize(n=\textcolor{black}{15})}} & Visits / encounters per day or week; appointment volume; clinic days & \textcolor{black}{9} & \cite{rotenstein2026changes,toussi2026impact,holmgren2026ambient,pearlman2025use,pelletier2025effect,husa2026ambient,vallabhaneni2026reduction,dutta2026medical,lawrence2026implementing} \\*
 & Completed sessions / caseloads / working hours & 2 & \cite{mccrudden2026ai,rui2025qualitative} \\*
 & Productivity \% / clinical-hour utilization & 1 & \cite{evans2025impact} \\*
 & Notes per month & 1 & \cite{wright2025effect} \\*
 & Self-established item: capacity to add patients & \textcolor{black}{2} & \cite{olson2025use,wojda2026burden} \\
\cmidrule(l{0pt}r{0pt}){1-4}
\multirow[t]{1}{=}{\raggedright wRVUs / productivity units \textit{\scriptsize(n=\textcolor{black}{7})}} & Work Relative Value Units per encounter, week, day, or month & \textcolor{black}{7} & \cite{haberle2024impact,shin2025influence,holmgren2026ambient,moura2026hybrid,waken2026longitudinal,husa2026ambient,dutta2026medical} \\
\cmidrule(l{0pt}r{0pt}){1-4}
\multirow[t]{4}{=}{\raggedright Revenue / billing / coding\newline\textit{\scriptsize(n=4)}} & Revenue per clinician & 1 & \cite{rotenstein2026changes} \\*
 & Fees per visit / day; diagnostic codes per visit; date-of-service to bill posting & 1 & \cite{toussi2026impact} \\*
 & Claim denial proportion (revenue integrity) & 1 & \cite{holmgren2026ambient} \\*
 & Billing / E\&M coding level & 1 & \cite{harvey2025ambient} \\
\cmidrule(l{0pt}r{0pt}){1-4}
\multirow[t]{1}{=}{\raggedright Tool cost / ROI \textit{\scriptsize(n=\textcolor{black}{2})}} & Direct cost of the scribe tool (e.g.\ vs.\ human scribes) & \textcolor{black}{2} & \cite{cao2024artificial,kapoor2026improving} \\
\cmidrule(l{0pt}r{0pt}){1-4}
\rowcolor{gray!18}
\multicolumn{4}{@{}l}{\textbf{I.~Safety \& clinical outcomes}\quad\textup{\scriptsize(2 studies)}}\\*
\multirow[t]{1}{=}{\raggedright Safety risks / bias (perceived)\newline\textit{\scriptsize(n=1)}} & Self-established item: clinician-reported perceived safety risks, inaccuracies, and bias & 1 & \cite{lukac2025ambient} \\
\cmidrule(l{0pt}r{0pt}){1-4}
\multirow[t]{1}{=}{\raggedright Clinical / health outcomes\newline\textit{\scriptsize(n=1)}} & Composite depression outcome (depression ICD-10 coding, behavioral-health \& antidepressant referrals) & 1 & \cite{castro2026psychiatric} \\
\cmidrule(l{0pt}r{0pt}){1-4}
\rowcolor{gray!18}
\multicolumn{4}{@{}l}{\textbf{J.~Qualitative concepts (open-text \& interview)}\quad\textup{\scriptsize(16 studies)}}\\*
\multirow[t]{2}{=}{\raggedright Barriers, concerns \& limitations\newline\textit{\scriptsize(n=15)}} & Open-text survey response & 10 & \cite{pelletier2025effect,ballantyne2025using,chowdhury2026comparing,duggan2025clinician,marquis2026ai,rui2025qualitative,mccrudden2026ai,talman2026ambient,memon2026performance, baldwin2025early}
\\*
 & Semi-structured interview & 7 & \cite{nguyen2023implementing,evans2025impact,lawrence2026implementing,alpert2025evaluating,van2026ambient, memon2026performance, baldwin2025early}
\\
\cmidrule(l{0pt}r{0pt}){1-4}
\multirow[t]{2}{=}{\raggedright Perceived benefits \& advantages\newline\textit{\scriptsize(n=13)}} & Open-text survey response & 8 & \cite{pelletier2025effect,ballantyne2025using,rui2025qualitative,mccrudden2026ai,duggan2025clinician,marquis2026ai,talman2026ambient,memon2026performance}\\*
 & Semi-structured interview & 6 & \cite{baldwin2025ambient,evans2025impact,lawrence2026implementing,alpert2025evaluating,van2026ambient,memon2026performance}\\
\cmidrule(l{0pt}r{0pt}){1-4}
\multirow[t]{2}{=}{\raggedright Workflow integration \& implementation\newline\textit{\scriptsize(n=6)}} & Semi-structured interview & 6 & \cite{nguyen2023implementing,evans2025impact,baldwin2025ambient,lawrence2026implementing,alpert2025evaluating,van2026ambient}\\
& & \\
\cmidrule(l{0pt}r{0pt}){1-4}
\multirow[t]{2}{=}{\raggedright Clinician--patient interaction / alliance\newline\textit{\scriptsize(n=5)}} & Open-text survey response & 2 & \cite{marquis2026ai,memon2026performance}\\*
 & Semi-structured interview & 3 & \cite{evans2025impact,alpert2025evaluating,van2026ambient} \\
\cmidrule(l{0pt}r{0pt}){1-4}
\multirow[t]{2}{=}{\raggedright Trust, privacy \& ethico-legal concerns\newline\textit{\scriptsize(n=3)}} & Open-text survey response & 1 & \cite{ballantyne2025using} \\*
 & Semi-structured interview & 2 & \cite{evans2025impact,van2026ambient}\\
% \cmidrule(l{0pt}r{0pt}){1-4}
\end{longtable}
\endgroup
 
\subsection{ Efficiency and time}
Efficiency and time is by far the most frequently evaluated dimension, measured in \textcolor{black}{44} of \textcolor{black}{54} studies. The most common metrics are to estimate documentation efficiency and time, \eg time spent in notes (\textcolor{black}{30} studies) and after-hours (``pajama'') time (\textcolor{black}{29} studies) and most often derived from EHR audit logs such as Epic Signal (\eg \cite{rotenstein2026changes,north2025ambient,tierney2025ambient,waken2026longitudinal}). Chart-closure timeliness and, less often, consultation or cycle time are measured to give appointment level estimations. Crucially, these ostensibly objective metrics are operationalized inconsistently: 
after-hours work is measured in \textcolor{black}{29} studies using five non-equivalent definitions as pajama time, time outside scheduled hours, work outside work, and time on unscheduled days, with varying temporal windows and derivations across studies (\eg \cite{shin2025influence,chowdhury2026comparing,alpert2025evaluating,rotenstein2026changes}). 
Such divergence likely reflects the fact that EHR systems expose different fields for internal or administrative purposes. This is reasonable in itself, but it obstructs scientific comparison, which requires a shared convention across studies rather than merely clear definitions within each. As a result, even an ostensibly standard measure such as ``time in notes,'' computed per note, per appointment, per day, or per week, cannot be reconciled across studies.

\subsection{Adoption and utilization}
Adoption and utilization are commonly reported (\textcolor{black}{19} studies), capturing how many clinicians use the scribe and for what share of notes or visits (\eg \cite{mccrudden2026ai,north2025ambient,preiksaitis2026ambient,wright2026enterprise}). A subset of studies further quantify the tool's contribution to the note itself by measuring the proportion of note characters generated by the AI versus typed by the clinician \cite{north2025ambient,cao2024artificial,shin2025influence,duggan2025clinician,alpert2025evaluating}. These metrics %
are useful for gauging real-world uptake, which is exactly what adoption and utilization are meant to measure; the limitation is not the metrics but that adoption is sometimes reported without linking it to the outcomes that motivate or demotivate use.

\subsection{Note characteristics, quality, and accuracy}
Note-level outcomes are measured in \textcolor{black}{24} studies, but unevenly. Note length---a readily extracted surface feature---is the most common (\textcolor{black}{14} studies), whereas rated quality (\textcolor{black}{11} studies) and, especially, accuracy or fidelity (\textcolor{black}{6} studies) are far less frequent. However, while it is the most-measured note-level outcome (\textcolor{black}{14} studies), note length has no agreed valence \cite{rule2021length}, and most studies offer no clear position to interpret note length change.
Quality is assessed with a mixture of the validated Physician Documentation Quality Instrument (PDQI-9) \cite{memon2026performance,kapoor2026improving}, clinician thumbs-up/down ratings \cite{mccrudden2026ai}, and bespoke Likert items (\eg \cite{rabbani2025ambient,baldwin2025ambient,cao2024artificial}). Accuracy is measured only sporadically and with disparate methods, including ROUGE overlap against the final note \cite{memon2026performance}, Dice text-similarity with a generation-error rate \cite{mccrudden2026ai}, clinician-reported inaccuracies or bias \cite{lukac2025ambient}, cosine similarity \cite{alcazar2026deployment,alcazar2026real}, and clinician edit distance \cite{akhlaghi2026evaluation}. 
Notably, note accuracy, the dimension most directly linked to patient safety, is among the least frequently and most inconsistently measured. Although reducing clinician burden was the primary rationale for adoption, note correctness bears directly on patient safety, and its thin measurement leaves a central risk of the technology largely unexamined.

\subsection{Cognitive load and documentation burden}
Cognitive load and perceived documentation burden are widely assessed (\textcolor{black}{20} studies), but through a mixture of validated task-load instruments and single-item bespoke measures. The NASA Task Load Index and its variants appear in several studies (\eg \cite{schneider2026impact,harvey2025ambient,alpert2025evaluating,olson2025use}), alongside the four-item Provider Task Load index \cite{lukac2025ambient} and numerous self-established mental-demand or burden items (\eg \cite{baldwin2025ambient,rabbani2025ambient,evans2025impact,shin2025influence}). The same underlying construct is thus captured with non-comparable scales, so that a reduction in ``burden'' in one study cannot be quantitatively reconciled, making it difficult to build cumulative knowledge or verify effects across settings.

\subsection{Clinician well-being, burnout, and turnover}

Clinician well-being and burnout are frequently evaluated (\textcolor{black}{19} studies), most often with single-item or short validated burnout measures such as the Mini-Z (\eg \cite{nguyen2023implementing,alpert2025evaluating,lukac2025ambient,olson2025use}). Other studies use the Copenhagen Burnout Inventory \cite{chowdhury2026comparing}, the Professional Fulfillment Index and its work-exhaustion subscale \cite{rui2025qualitative,shah2025ambient}, or bespoke fatigue items \cite{shin2025influence,wright2026enterprise,harvey2026ambient,arko2026pre}. A smaller number extend to self-reported turnover intention \cite{shin2025influence,schneider2026impact,wright2026enterprise}. As with cognitive load, studies relied on several different burnout instruments each with its own scale, scoring, and cut-points, which means a burnout reduction reported in one study cannot be quantitatively combined with, or even placed on the same scale as, one from another.

\subsection{Usability and acceptability}

Usability and acceptability are assessed in \textcolor{black}{19} studies using a combination of validated scales, implementation-outcome frameworks, and bespoke items. The System Usability Scale \cite{shah2025ambient,duggan2025clinician, wright2025effect,kapoor2026improving}  and the Intervention Usability Scale \cite{nguyen2023implementing} appear alongside implementation frameworks such as the Weiner acceptability, feasibility, and appropriateness measures \cite{nguyen2023implementing}, the Proctor outcomes framework \cite{baldwin2025ambient}, and an adapted Unified Theory of Acceptance and Use of Technology \cite{shah2025ambient}. Likelihood-to-recommend and Net Promoter Scores recur as lightweight proxies for clinician acceptance (\eg \cite{mccrudden2026ai,duggan2025clinician,rui2025qualitative,pelletier2025effect}).

\subsection{Patient experience}
Patient experience is measured in \textcolor{black}{20} studies but shallowly. It is dominated by single-item, self-established questions---whether the clinician appeared attentive \cite{tan2026impact,harvey2025ambient,evans2025impact,rabbani2025ambient,harvey2026ambient}, maintained eye contact \cite{tan2026impact}, or spent enough time with the patient \cite{memon2026performance,patel2026effect,schneider2026impact}. Patient experience was also measured from clinician perspectives through self-report \cite{tierney2025ambient, olson2025use}. Validated patient-experience instruments are rare (Press Ganey and likelihood-to-recommend in one study \cite{haberle2024impact}) and Net promoter score \cite{rabbani2025ambient,alcazar2026deployment}, as is any measure of patient trust in AI \cite{evans2025impact,pelletier2025effect} and their comprehension of care plan \cite{pelletier2025effect,olson2025use,wojda2026burden}. The patient's perspective, despite being directly affected by ambient recording, is therefore captured mostly through ad hoc items that vary from study to study.
Only one study used the patient opt-out rate \cite{haberle2024impact}, a behavioral instrument, to measure patient consent.

\subsection{Productivity, billing, and financial outcomes}
Operational and financial outcomes are reported in \textcolor{black}{22} studies, typically as secondary endpoints: visit or encounter volume (\eg \cite{rotenstein2026changes,holmgren2026ambient,pearlman2025use,pelletier2025effect,olson2025use,evans2025impact}), work relative value units \cite{haberle2024impact,shin2025influence,holmgren2026ambient,moura2026hybrid,waken2026longitudinal,husa2026ambient,dutta2026medical}, billing and coding levels \cite{rotenstein2026changes,holmgren2026ambient,toussi2026impact,harvey2025ambient}, and, rarely, direct cost \cite{cao2024artificial,kapoor2026improving}. Direct cost-effectiveness or return-on-investment analyses are largely absent, even though the business case is central to enterprise adoption.

\subsection{Safety and clinical outcomes}

Safety and downstream clinical outcomes are the least-measured dimension, appearing in only two studies \cite{lukac2025ambient,castro2026psychiatric}. One captures clinician-perceived safety risks, inaccuracies, and bias \cite{lukac2025ambient}; the other estimates a composite clinical outcome from documentation \cite{castro2026psychiatric}. No study systematically audited scribe-generated notes for clinically significant errors or omissions against a reference standard. For a technology that drafts the medical record, the near-absence of direct safety evaluation is striking. 
Understandably, early evaluations have focused on the primary drivers of adoption, clinician burden and burnout, as well as readily available, short-term, EHR-derived metrics. We also observed that only one of the 14 studies spanning over a year or more look at safety risks associated with AI scribes that too from clinicians perspectives only. Fully understanding the impact of ambient scribes in healthcare, however, will require a broader set of metrics spanning safety and clinical outcomes and multiple time horizons, from immediate documentation effects to durable effects on care.

\subsection{Qualitative concepts}
A subset of studies used open-text survey responses and semi-structured interviews to surface concepts that structured metrics miss (\textcolor{black}{16} studies). These qualitative data recurrently raise barriers, concerns, and limitations \cite{pelletier2025effect,ballantyne2025using,chowdhury2026comparing,duggan2025clinician,marquis2026ai,rui2025qualitative,mccrudden2026ai,nguyen2023implementing,evans2025impact, talman2026ambient,lawrence2026implementing, baldwin2025early,alpert2025evaluating,van2026ambient,memon2026performance}; perceived benefits and advantages \cite{talman2026ambient,lawrence2026implementing, pelletier2025effect,ballantyne2025using,rui2025qualitative,mccrudden2026ai,duggan2025clinician,marquis2026ai,baldwin2025ambient,evans2025impact, memon2026performance,alpert2025evaluating,van2026ambient}; workflow integration and implementation experience \cite{nguyen2023implementing,evans2025impact,baldwin2025ambient, lawrence2026implementing,alpert2025evaluating, van2026ambient}; effects on the clinician--patient interaction \cite{marquis2026ai,evans2025impact, memon2026performance,alpert2025evaluating,van2026ambient}; and trust, privacy, and ethico-legal concerns \cite{ballantyne2025using,evans2025impact, van2026ambient}. Notably, accuracy \cite{ballantyne2025using,marquis2026ai,chowdhury2026comparing}, editing burden \cite{marquis2026ai,duggan2025clinician}, and medico-legal risk \cite{ballantyne2025using,evans2025impact} appear in free text even though they are seldom captured quantitatively, suggesting that clinicians are attuned to risks the formal metrics do not track.

\begin{figure}[t]\centering
\includegraphics[width=0.7\textwidth]{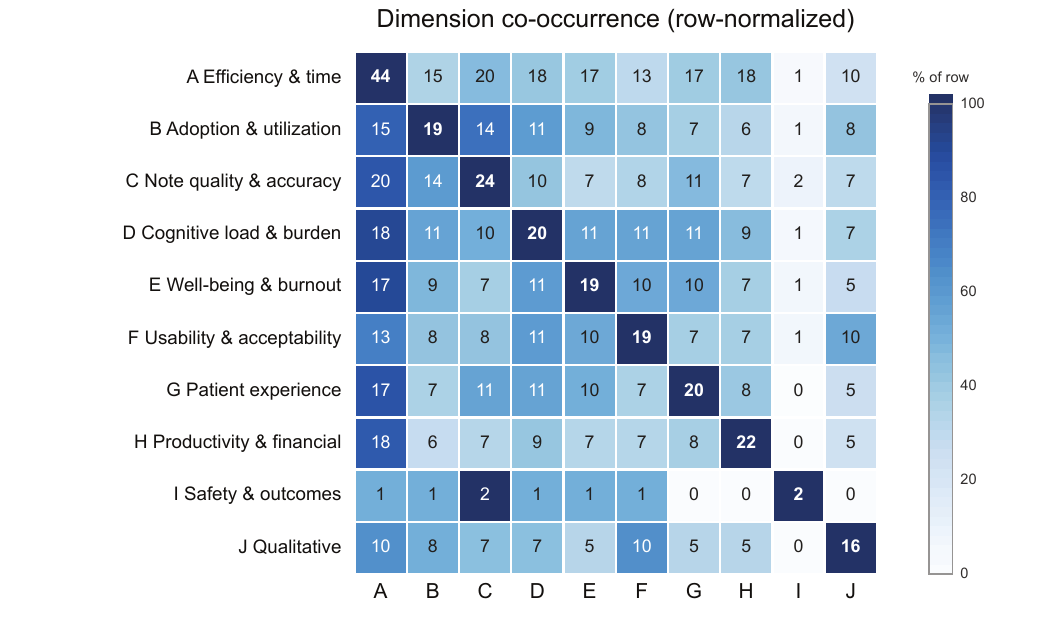}
\caption{%\cmh{this figure does not make sense to me. for example , top left corner, why ``10'' are coded differently?}  
Pairwise co-occurrence of evaluation dimensions across the \textcolor{black}{54} studies. Each cell reports the number of studies measuring both the row and column dimension (the diagonal gives the number measuring that dimension). Cell shading encodes co-occurrence as a percentage of the row dimension's studies (0--100\%); it is therefore read row-wise and is not symmetric \am{(each row is normalized by its own total, two cells with the same count can appear in different shades)}. Documentation efficiency (A) co-occurs with nearly every other dimension, whereas safety (I), measured in only two studies, rarely co-occurs with any. }\label{fig:cooc}
\label{fig:cooc}
\end{figure}

\section{Results: The Evaluation Landscape}\label{sec:landscape}
Stepping back from individual dimensions to the evidence base as a whole reveals two patterns: measurement is  spread across multiple dimensions yet fragmented within them, and studies rarely measure the same things in the same ways, so their results do not line up. 
\subsection{Concentration and non-comparability of measurement}
Compiling the ten dimensions together shows that measurement is  measurement is spread unevenly across many dimensions and ignores critical ones; documentation efficiency is evaluated in \textcolor{black}{44} of \textcolor{black}{54} studies, followed by \textcolor{black}{note characteristics (24)}, \textcolor{black}{productivity (22)}, and \textcolor{black}{cognitive load (20)}, whereas safety and clinical outcomes are evaluated in only two (Figure~\ref{fig:map}). 
The dimensions most closely tied to patient welfare, including safety, patient experience in depth, and downstream clinical effects, are among the least studied. Consequently, the evidence base is strongest where measurement is easiest and weakest where the stakes are highest.
 
This concentration coexists with pervasive fragmentation \emph{within} dimensions. As Figure~\ref{fig:map} 
and the instrument column of Table~\ref{tab:metrics} show, a single construct is typically measured in several non-equivalent ways: across the \textcolor{black}{54} studies we identified \textcolor{black}{96} distinct operationalizations, of which 35 (36\%) are used by only one study and just 29 by five or more. Documentation efficiency alone is measured with \textcolor{black}{18} different operationalizations. Even seemingly identical metrics diverge: ``after-hours work'' appears as pajama time, time outside scheduled hours, work outside work, or time on unscheduled days, and ``time in notes'' is normalized per note, per appointment, per day, or per week. The same construct is thus rarely measured the same way twice, which prevents combining results even when studies claim to address the same outcome.
 
\subsection{Within-study breadth and metric co-occurrence}

Individual studies are internally coherent but collectively non-comparable. Studies measured a median of four of the ten dimensions (range 1--\textcolor{black}{8}) and six distinct metrics (range 2--13); \textcolor{black}{nineteen} addressed five or more dimensions, whereas \textcolor{black}{five} measured only one. The row-normalized view in Figure~\ref{fig:cooc} exposes a pervasive \emph{asymmetry} in how dimensions are measured together. Documentation efficiency was measured alongside nearly every other dimension; a large majority of the studies that measured cognitive load (n=\textcolor{black}{18}, \textcolor{black}{90\%}), well-being (n=\textcolor{black}{17}, \textcolor{black}{89\%}), patient experience (n=\textcolor{black}{17}, \textcolor{black}{85\%}), \textcolor{black}{note quality (n=20, 83\%)}, \textcolor{black}{productivity (n=18, 82\%)}, or adoption (n=\textcolor{black}{15}, \textcolor{black}{79\%}) also measured efficiency, yet the reverse did not hold---among the \textcolor{black}{44} studies that measured efficiency, no other dimension was measured by even half (at most 45\%, for \textcolor{black}{note characteristics}). In other words, most studies that measured a given construct also measured efficiency, but among the studies that measured efficiency, most did not measure that construct.
The same one-directional pattern appears elsewhere: both studies that measured safety also measured note quality, whereas only a small minority of note-quality studies (\textcolor{black}{2 of 24}) measured safety. Pairings that did not involve efficiency were comparatively uncommon overall. No standard bundle of outcomes emerged even within the most-measured dimension; the most frequent combination of efficiency sub-metrics (time in notes, after-hours work, and chart-closure timeliness) appeared in only \textcolor{black}{nine} of the \textcolor{black}{44} studies measuring efficiency. Considered as whole measurement profiles rather than pairwise overlaps, the \textcolor{black}{54} studies span \textcolor{black}{40} distinct dimension-combinations (Figure~\ref{fig:map}). Even the single most common combination---covering only efficiency and productivity (dimensions A and H)---appears in only \textcolor{black}{five} studies \textcolor{black}{\cite{husa2026ambient,moura2026hybrid,pearlman2025use,rotenstein2026changes,vallabhaneni2026reduction}}, and \textcolor{black}{every other combination is shared by at most three studies}. In practice, then, there is no shared measurement template: two studies drawn at random %, with same research question or problem, 
almost never evaluate the same set of dimensions, so they rarely share even an overall footprint. The problem is that even when two studies share the same focus, differences in operationalization keep their results from being compared.

The same pattern holds at the instrument level: among studies measuring after-hours work, \textcolor{black}{twenty-two} relied on a single operationalization and only \textcolor{black}{seven} combined two or more, and burnout and cognitive load are each split across four different instruments, with the most common instrument used in only \textcolor{black}{7 of 13} and \textcolor{black}{10 of 14} studies, respectively.
Consequently, even studies that appear comprehensive typically measured each dimension through a single metric and operationalization of their own choosing, and these choices differ from one study to the next, so that thoroughness within a study does not translate into comparability across studies.

\section{Discussion}\label{sec:discussion}
 
Evaluation of ambient AI scribes has not kept pace with their deployment. Many health systems have obtained executive support for pilot programs without first establishing the data-analytic and measurement infrastructure needed to assess impact rigorously across process, experience, financial, and quality outcomes; as a result, decision-makers often lack the evidence required to justify scaling and authorize continued investment only up to limited thresholds of spend or scale \cite{peterson2025adoption}. Comparison across organizations is further complicated by differences in operations and in how metrics are defined, and health-system leaders themselves have identified a need for standardized metric definitions, aligned outcome domains, and more rigorous study designs to enable cross-site comparison \cite{peterson2025adoption}.
 
These gaps reflect a broader absence of standards for evaluating clinical AI. Evaluation frameworks are widely expected to be comprehensive, standardized, repeatable, and transparent \cite{wells2025practical}, yet established approaches for both pre-implementation review and post-implementation monitoring of healthcare AI remain limited. Existing governance instruments offer only partial guidance: the EU AI Act is legally binding but has been criticized for an unclear and inflexible definition of ``high-risk'' AI that fits healthcare poorly, where risk depends on the specific tool, the degree of human oversight, and the clinical use case; and in the United States, the FDA's Software-as-a-Medical-Device guidance \cite{fda_samd}, the NIST AI Risk Management Framework \cite{nist_ai_rmf}, and the Blueprint for an AI Bill of Rights \cite{ostp_ai_bill_of_rights} are non-binding and provide little concrete direction for implementation within real-world health systems \cite{wells2025practical}.
 
Scribe-specific evaluation efforts have begun to emerge, but each addresses only part of the problem. One framework assesses transcription, diarization, and medical-note generation against criteria such as fluency, completeness, and factuality \cite{wang2025evaluation}; while valuable for characterizing model performance, it does not capture the benefits realized once a tool is deployed within an organization, nor the quality of the resulting human-AI collaboration. Another framework, FAIR-AI, structures governance as a two-stage process with an initial low-risk screening pathway followed by in-depth review for solutions that do not pass the screen, but it is oriented toward risk assessment rather than the choice and definition of outcome metrics \cite{wells2025practical}. Work on accuracy remains limited in AI scribe real-world deployments. Existing evaluations to date find that no system is error-free \cite{ha2025evaluating}, and only recently has an instrument been developed and validated specifically to assess the accuracy and safety of scribe outputs \cite{biro2025accuracy}, which is a prerequisite for comparing systems or tracking performance over time.

Our review contributes an empirical map of what is actually measured. The field has converged on a narrow set of data sources, primarily relying on surveys and audit logs (Section~\ref{sec:datasources}). However, despite this reliance on common data sources, measurement remains inconsistent, with the same constructs operationalized in many different ways (Section~\ref{sec:landscape}). The limitations below follow directly from this combination.

\subsection{Limitations of current evaluation practice}
 Despite  growing body of work on AI scribes, current evaluation practice has several limitations.
\subsubsection{Study designs and methods}
Current evaluations are dominated by short, single-site, and largely uncontrolled designs that are vulnerable to novelty effects, secular trends, and confounding. Rigor and duration are, moreover, inversely related in this literature: the three randomized trials all ran only three to four months \cite{shin2025influence,lukac2025ambient,chowdhury2026comparing}, whereas most longer studies are retrospective or observational, so no design combines a concurrent control group with follow-up beyond a single quarter.
 
Compounding these design weaknesses is heavy reliance on clinician self-report as the primary lens. Self-report is valuable and often the only practical way to capture experience, but when it is the main or sole source it exposes findings to social-desirability and expectation bias.
Four of the ten evaluation dimensions---cognitive load, clinician well-being, usability, and patient experience---are measured entirely or almost entirely by self-report (validated or self-established). The dimensions that carry the technology's core value proposition, namely reduced burden, reduced burnout, and acceptability, are therefore precisely those that lack objective corroboration. 

Methodological rigor is allocated to the lowest-stakes outcomes. None of the three randomized trials 
%\cite{olson2025use,haberle2024impact,pelletier2025effect} 
\cite{shin2025influence,lukac2025ambient,chowdhury2026comparing}
measure patient experience. The only safety measure among them is a bespoke clinician-reported item. Randomization has been spent almost entirely on clinician-experience endpoints. Furthermore, these trials cannot be blinded, as clinicians know whether they are dictating to a scribe. As a result, randomization does not protect self-reported burnout, burden, or usability endpoints from expectation effects. Randomization is most valuable for objective endpoints, which are the endpoints current trials under-emphasize.

\subsubsection{Metrics and operationalization}
The fragmentation documented in Section~\ref{sec:landscape} has a direct consequence for cumulative evidence: because the same construct is rarely captured with the same instrument, definition, or time window twice, effect estimates cannot be placed on a common scale, and the literature cannot converge toward consensus values even when studies  address the same outcome.
 
This obstacle is compounded by incomplete reporting of exposure. Only \textcolor{black}{15} studies report both a documentation-efficiency outcome and adoption or utilization, and \textcolor{black}{21} report a change in documentation time without reporting what proportion of clinicians or notes actually used the tool. Effects are thus implicitly conditioned on continued use: clinicians who find the tool unhelpful tend to stop using it, so per-user estimates are inflated by survivorship. Only five studies report adoption concentration, and although five quantify the AI's contribution to the note (the share of characters AI-generated versus clinician-typed, which is a natural continuous measure of exposure), none relate this exposure to the magnitude of any outcome, leaving the field without exposure-response evidence.
 
Finally, several widely used metrics lack a stated direction of benefit. Note length is the most-measured note-level outcome, yet it has no agreed valence \cite{rule2021length}; longer notes may reflect greater completeness or unnecessary verbosity, and because most studies take no interpretive position, the metric is unusable for decision-making regardless of how consistently it is measured.
 
\subsubsection{Dimensions absent from the evidence base}
Beyond the dimensions that are measured but only thinly (Section~\ref{sec:landscape}), several are essentially absent from the evidence base altogether. We found no evaluation of equity, including whether benefits and harms are distributed evenly across patient populations or whether accuracy degrades for particular accents, languages, or demographic groups. Note comprehensibility to patients, increasingly relevant as notes are shared through patient portals, is likewise underexamined  (n=\textcolor{black}{3}, \cite{olson2025use,pelletier2025effect,wojda2026burden}), as are downstream effects on the clinicians and care-team members who later read scribe-generated notes. No study addresses potential deskilling of documentation or clinical-reasoning abilities under prolonged reliance on automation, and none characterizes the tools' failure modes, that is, how, when, and with what consequences they err. Privacy, security, and data governance, though central to the ambient capture of clinical conversations, surface mainly in passing in free-text comments or interviews \cite{ballantyne2025using,evans2025impact,van2026ambient} rather than being evaluated directly. Together these omissions define a concrete agenda for future evaluation.
 
\subsection{Open problems and opportunities}
The limitations in current practice point to several open problems and opportunities for improving how ambient AI scribes are evaluated.
\subsubsection{Toward a standardized, multi-dimensional core metric set}
Building comparable and cumulative evidence requires two elements: a small core set of metrics reported in every evaluation, ensuring that safety and patient experience are not overlooked simply because they are harder to measure, and the use of shared, validated instruments beyond this core whenever studies pursue the same evaluation goal, allowing results measuring the same construct to be compared.
When validated instruments exist, they should be preferred and explicitly identified, for example, reporting by a Mini-Z or NASA-TLX score rather than an ad hoc measure, to enable comparisons on a common scale. Health-system leaders have called for this kind of standardization, including agreed-upon metric definitions and aligned outcome domains, to enable comparison across sites \cite{peterson2025adoption}.
 
\subsubsection{Measuring accuracy, safety, and downstream effects}
Note accuracy, the dimension most directly linked to patient safety, is among the least frequently and most inconsistently measured. The field needs direct, standardized assessment of note accuracy and safety, including structured audits of scribe-generated notes for clinically significant errors and omissions against a reference standard, rather than reliance on clinician impressions; recently developed accuracy-and-safety instruments offer a foundation for this \cite{biro2025accuracy,wang2025evaluation}. Evaluation should likewise extend to the downstream and ripple effects that current studies omit, including orders and referrals generated, time spent by downstream note readers, equitable performance across populations, and cost-effectiveness.
 
\subsubsection{Standardizing ``objective'' EHR metrics}
Even the ostensibly objective EHR metrics require standardization. Epic Signal time-in-notes, for example, is variously reported per note, per appointment, per day, or as a weekly average, and these normalizations can move in opposite directions from the same underlying effect: if a scribe increases visit volume, time-in-notes per day can rise even as time-in-notes per note falls. A substantial share of the apparent disagreement in this literature may therefore be a normalization artifact rather than genuine heterogeneity, arising from differences in the denominator across studies rather than true differences in effect. Harmonized definitions of time-in-notes and after-hours work would make the field's most common metrics comparable across studies.
 
\subsubsection{Stronger designs and transparent reporting}
More rigorous study designs, such as multi-site, controlled (when feasible), and longitudinal studies, are needed to distinguish sustained effects from novelty effects. Moreover, as argued above, randomized designs should be directed at objective endpoints, where they add the most value. Transparent reporting of data sources, operationalizations, instruments used, and exposure (adoption and utilization) would make heterogeneous studies interpretable and, ideally, can be compared or combined across other studies to build collective knowledge. Established evaluation frameworks from implementation science, health IT, and natural language processing \cite{wang2025evaluation,wells2025practical} provide a starting point that can be adapted for this purpose.
 
\subsection{Limitations of this review}
This review has several limitations. The literature is moving quickly and our search has a fixed cutoff, so relevant work and preprints may have been missed. Given the heterogeneity of outcomes, we synthesized findings through structured mapping and vote-counting rather than meta-analysis, which does not weight studies by size or quality. 
Finally, coding metrics to dimensions and instruments involves judgment; although one researcher performed the coding and a second reviewer independently verified it with full agreement, some classifications remain interpretive.

\section{Conclusion}\label{sec:conclusion}
 
Evaluation of ambient AI scribes in healthcare is expanding quickly, but it remains narrowly concentrated, self-report-heavy, and inconsistently operationalized. The field measures documentation efficiency thoroughly, while note accuracy and safety are measured rarely and patient experience mostly through ad hoc items. Even when studies address the same construct, they use different instruments and definitions, so their results cannot be combined. Moving toward a standardized, holistic, and multi-dimensional evaluation approach, grounded in validated instruments and standardized definitions, is essential to produce the comparable, cumulative evidence needed to guide the safe and effective scaling of these tools.

%\newpage
\section*{Statement of AI use}
AI tools were used for copy-editing, ideation and design of visualizations, and double-checking counts for analysis. All AI-assisted outputs were manually reviewed and verified by the authors.

\section*{Acknowledgment}
This work was performed under the following financial assistance award 60NANB25D119 from U.S. Department of Commerce, National Institute of Standards and Technology (NIST). We thank Michael Enaye for assistance with searching and identifying papers for inclusion in the review.

\section*{Author CRediT} A.M.: Conceptualization, Methodology, Validation, Formal analysis, Investigation, Data curation,
Writing—Original draft, Writing—Review and editing, Visualization, Project Administration. S.C.: Conceptualization, Methodology, Validation, Formal analysis, Investigation, Data curation,
Writing—Original draft, Writing—Review and editing, Visualization. 
L.F.: Conceptualization, Methodology, Validation, Formal analysis, Investigation, Data curation,
Writing—Original draft, Writing—Review and editing. J.J.: Conceptualization, Methodology, Formal analysis, Investigation, Data curation,
Writing—Original draft. T.J.: Conceptualization, Writing—Review and editing. Y.C.: Conceptualization, Writing—Review and editing. K.G.: Conceptualization, Writing—Review and editing. C.H.: Conceptualization, Methodology, Writing—Review and editing, Supervision, Funding acquisition. 

\section*{Disclaimer}
Certain equipment, instruments, software, or materials are identified in this paper in order to specify the experimental procedure adequately. Such identification is not intended to imply recommendation or endorsement of any product or service by NIST, nor is it intended to imply that the materials or equipment identified are necessarily the best available for the purpose.

%\appendix
\newpage

%\clearpage
\bibliographystyle{ACM-Reference-Format}
\bibliography{references}


\begin{thebibliography}{79}

%%% the Web address.

\ifx \showCODEN    \undefined \def \showCODEN     #1{\unskip}     \fi
\ifx \showISBNx    \undefined \def \showISBNx     #1{\unskip}     \fi
\ifx \showISBNxiii \undefined \def \showISBNxiii  #1{\unskip}     \fi
\ifx \showISSN     \undefined \def \showISSN      #1{\unskip}     \fi
\ifx \showLCCN     \undefined \def \showLCCN      #1{\unskip}     \fi
\ifx \shownote     \undefined \def \shownote      #1{#1}          \fi
\ifx \showarticletitle \undefined \def \showarticletitle #1{#1}   \fi
\ifx \showURL      \undefined \def \showURL       {\relax}        \fi
% invisible to TeX
\providecommand\bibfield[2]{#2}
\providecommand\bibinfo[2]{#2}
\providecommand\natexlab[1]{#1}
\providecommand\showeprint[2][]{arXiv:#2}

\bibitem[Adler-Milstein et~al\mbox{.}(2026)]%
        {adler2026subjective}
\bibfield{author}{\bibinfo{person}{Julia Adler-Milstein}, \bibinfo{person}{Orianna DeMasi}, \bibinfo{person}{Hossein Soleimani}, \bibinfo{person}{Sarah Beck}, \bibinfo{person}{Maria~E Byron}, \bibinfo{person}{Aris Oates}, \bibinfo{person}{Robert Thombley}, \bibinfo{person}{Jinoos Yazdany}, {and} \bibinfo{person}{Sara~G Murray}.} \bibinfo{year}{2026}\natexlab{}.
\newblock \showarticletitle{Subjective and Objective Impacts of Ambulatory AI Scribes.}
\newblock \bibinfo{journal}{\emph{American Journal of Managed Care}} \bibinfo{volume}{32}, \bibinfo{number}{1} (\bibinfo{year}{2026}).
\newblock


\bibitem[Akhlaghi et~al\mbox{.}(2026)]%
        {akhlaghi2026evaluation}
\bibfield{author}{\bibinfo{person}{Hamed Akhlaghi}, \bibinfo{person}{Sam Freeman}, \bibinfo{person}{Kevin Sun}, \bibinfo{person}{Neo Nie}, \bibinfo{person}{John Ding}, \bibinfo{person}{Leo Chen}, \bibinfo{person}{Eden Pham}, \bibinfo{person}{Brendan Morrissey}, {and} \bibinfo{person}{Jonathan Karro}.} \bibinfo{year}{2026}\natexlab{}.
\newblock \showarticletitle{Evaluation of an AI Scribe Tool in the Emergency Department: A Single-Arm Observational Study}.
\newblock \bibinfo{journal}{\emph{Emergency Medicine Australasia}} \bibinfo{volume}{38}, \bibinfo{number}{3} (\bibinfo{year}{2026}), \bibinfo{pages}{e70272}.
\newblock


\bibitem[Alc{\'a}zar-Peral et~al\mbox{.}(2026a)]%
        {alcazar2026real}
\bibfield{author}{\bibinfo{person}{Jose~Mar{\'\i}a Alc{\'a}zar-Peral}, \bibinfo{person}{Juan~Antonio {\'A}lvaro-de~la Parra}, \bibinfo{person}{Daniel Blanco}, \bibinfo{person}{{\'A}ngel Blanco}, \bibinfo{person}{Mar{\'\i}a~Elvira Barrios}, \bibinfo{person}{Ion Crist{\'o}bal}, {and} \bibinfo{person}{Cristina Caram{\'e}s}.} \bibinfo{year}{2026}\natexlab{a}.
\newblock \showarticletitle{Real-world evaluation of an ambient AI scribe in Spanish outpatient care after 2.3 million uses: impact on clinician experience, semantic agreement, and workflow efficiency}.
\newblock \bibinfo{journal}{\emph{Frontiers in Digital Health}}  \bibinfo{volume}{8} (\bibinfo{year}{2026}), \bibinfo{pages}{1874919}.
\newblock


\bibitem[Alc{\'a}zar-Peral et~al\mbox{.}(2026b)]%
        {alcazar2026deployment}
\bibfield{author}{\bibinfo{person}{Jose~Mar{\'\i}a Alc{\'a}zar-Peral}, \bibinfo{person}{Juan~Antonio {\'A}lvaro-de~la Parra}, \bibinfo{person}{Paolo Ciardo}, \bibinfo{person}{Henry~Jos{\'e} Marcano}, \bibinfo{person}{Miguel Calder{\'o}n}, \bibinfo{person}{Carlos Villar}, \bibinfo{person}{Daniel Blanco}, \bibinfo{person}{Ion Crist{\'o}bal}, {and} \bibinfo{person}{Cristina Caram{\'e}s}.} \bibinfo{year}{2026}\natexlab{b}.
\newblock \showarticletitle{Deployment of an ambient AI scribe in emergency care: A 12-month evaluation in a large Spanish hospital network}.
\newblock \bibinfo{journal}{\emph{International Journal of Medical Informatics}} (\bibinfo{year}{2026}), \bibinfo{pages}{106584}.
\newblock


\bibitem[Alpert et~al\mbox{.}(2025)]%
        {alpert2025evaluating}
\bibfield{author}{\bibinfo{person}{Jordan~M Alpert}, \bibinfo{person}{Robert Saper}, \bibinfo{person}{Eric Boose}, \bibinfo{person}{Jessica Ruff}, \bibinfo{person}{Kevin Hopkins}, \bibinfo{person}{Devyn Gaskins}, \bibinfo{person}{Ning Guo}, \bibinfo{person}{Jakob~P Schneider}, \bibinfo{person}{Aryn Giffi~Scibona}, \bibinfo{person}{James Gutierrez}, {et~al\mbox{.}}} \bibinfo{year}{2025}\natexlab{}.
\newblock \showarticletitle{Evaluating an artificial intelligence scribe for clinical documentation}.
\newblock \bibinfo{journal}{\emph{Digital Health}}  \bibinfo{volume}{11} (\bibinfo{year}{2025}), \bibinfo{pages}{20552076251395588}.
\newblock


\bibitem[Arko et~al\mbox{.}(2026)]%
        {arko2026pre}
\bibfield{author}{\bibinfo{person}{Leopold Arko}, \bibinfo{person}{Tony~Huy Nguyen}, \bibinfo{person}{Michelle Stoffel}, \bibinfo{person}{Jaya Kumar}, \bibinfo{person}{Sameer Badlani}, \bibinfo{person}{Genevieve~B Melton}, {and} \bibinfo{person}{Rebecca Markowitz}.} \bibinfo{year}{2026}\natexlab{}.
\newblock \showarticletitle{Pre--Post Evaluation of Documentation Burden, Time Perception, and Cognitive Workload of Ambient AI Scribe Tools}.
\newblock \bibinfo{journal}{\emph{AMIA Summits on Translational Science Proceedings}}  \bibinfo{volume}{2026} (\bibinfo{year}{2026}), \bibinfo{pages}{202}.
\newblock


\bibitem[Atiku et~al\mbox{.}(2026)]%
        {atiku2026usability}
\bibfield{author}{\bibinfo{person}{Samuel Atiku}, \bibinfo{person}{Olufisayo Olakotan}, {and} \bibinfo{person}{Kehinde Owolanke}.} \bibinfo{year}{2026}\natexlab{}.
\newblock \showarticletitle{Usability-Related Barriers and Facilitators Influencing the Adoption and Use of AI Scribes in Healthcare: A Scoping Review}.
\newblock \bibinfo{journal}{\emph{Journal of Evaluation in Clinical Practice}} \bibinfo{volume}{32}, \bibinfo{number}{1} (\bibinfo{year}{2026}), \bibinfo{pages}{e70365}.
\newblock


\bibitem[Baldwin et~al\mbox{.}(2025)]%
        {baldwin2025early}
\bibfield{author}{\bibinfo{person}{Abigail Baldwin}, \bibinfo{person}{Allison Lipitz-Snyderman}, \bibinfo{person}{Suryan Goyal}, \bibinfo{person}{Gilad Kuperman}, \bibinfo{person}{Peter~D Stetson}, \bibinfo{person}{Susan Chimonas}, \bibinfo{person}{Adam Watson}, {and} \bibinfo{person}{Maclain Damstrom}.} \bibinfo{year}{2025}\natexlab{}.
\newblock Early insights into the impact of an ambient AI scribe solution on clinical documentation to reduce clinician burnout in oncology.
\newblock


\bibitem[Baldwin-Medsker et~al\mbox{.}(2025)]%
        {baldwin2025ambient}
\bibfield{author}{\bibinfo{person}{Abigail~E Baldwin-Medsker}, \bibinfo{person}{Susan Chimonas}, \bibinfo{person}{Suryan~P Goyal}, \bibinfo{person}{Adam~J Watson}, \bibinfo{person}{Maclain~R Damstrom}, \bibinfo{person}{Gilad~J Kuperman}, \bibinfo{person}{Tiffany Traina}, \bibinfo{person}{Benjamin~R Roman}, \bibinfo{person}{Peter~D Stetson}, {and} \bibinfo{person}{Allison Lipitz-Snyderman}.} \bibinfo{year}{2025}\natexlab{}.
\newblock \showarticletitle{Ambient Artificial Intelligence Scribes in Oncology: Adoption, Feasibility, Acceptability, and Appropriateness}.
\newblock \bibinfo{journal}{\emph{Applied Clinical Informatics}} \bibinfo{volume}{16}, \bibinfo{number}{04} (\bibinfo{year}{2025}), \bibinfo{pages}{995--1004}.
\newblock


\bibitem[Ballantyne et~al\mbox{.}(2025)]%
        {ballantyne2025using}
\bibfield{author}{\bibinfo{person}{Angela Ballantyne}, \bibinfo{person}{Rochelle Style}, \bibinfo{person}{Maria Stubbe}, \bibinfo{person}{Samantha Murton}, {and} \bibinfo{person}{Tony Dowell}.} \bibinfo{year}{2025}\natexlab{}.
\newblock \showarticletitle{Using AI scribes in New Zealand primary care consultations: an exploratory survey}.
\newblock \bibinfo{journal}{\emph{The Journal of Primary Health Care}} (\bibinfo{year}{2025}).
\newblock


\bibitem[Biro et~al\mbox{.}(2025)]%
        {biro2025accuracy}
\bibfield{author}{\bibinfo{person}{Joshua Biro}, \bibinfo{person}{Jessica~L Handley}, \bibinfo{person}{Nathan~K Cobb}, \bibinfo{person}{Varsha Kottamasu}, \bibinfo{person}{Jeffrey Collins}, \bibinfo{person}{Seth Krevat}, {and} \bibinfo{person}{Raj~M Ratwani}.} \bibinfo{year}{2025}\natexlab{}.
\newblock \showarticletitle{Accuracy and safety of AI-enabled scribe technology: instrument validation study}.
\newblock \bibinfo{journal}{\emph{Journal of Medical Internet Research}}  \bibinfo{volume}{27} (\bibinfo{year}{2025}), \bibinfo{pages}{e64993}.
\newblock


\bibitem[Cao et~al\mbox{.}(2024)]%
        {cao2024artificial}
\bibfield{author}{\bibinfo{person}{David~Y Cao}, \bibinfo{person}{Jamie~R Silkey}, \bibinfo{person}{Michael~C Decker}, {and} \bibinfo{person}{Karolyn~A Wanat}.} \bibinfo{year}{2024}\natexlab{}.
\newblock \showarticletitle{Artificial intelligence-driven digital scribes in clinical documentation: pilot study assessing the impact on dermatologist workflow and patient encounters}.
\newblock \bibinfo{journal}{\emph{JAAD international}}  \bibinfo{volume}{15} (\bibinfo{year}{2024}), \bibinfo{pages}{149--151}.
\newblock


\bibitem[Carayon et~al\mbox{.}(2006)]%
        {carayon2006work}
\bibfield{author}{\bibinfo{person}{PASH Carayon}, \bibinfo{person}{A~Schoofs Hundt}, \bibinfo{person}{Ben-Tzion Karsh}, \bibinfo{person}{Ayse~P Gurses}, \bibinfo{person}{Carla~J Alvarado}, \bibinfo{person}{Mark Smith}, {and} \bibinfo{person}{P~Flatley Brennan}.} \bibinfo{year}{2006}\natexlab{}.
\newblock \showarticletitle{Work system design for patient safety: the SEIPS model}.
\newblock \bibinfo{journal}{\emph{BMJ Quality \& Safety}} \bibinfo{volume}{15}, \bibinfo{number}{suppl 1} (\bibinfo{year}{2006}), \bibinfo{pages}{i50--i58}.
\newblock


\bibitem[Carmichael et~al\mbox{.}(2026)]%
        {carmichael2026artificial}
\bibfield{author}{\bibinfo{person}{Gavin~J Carmichael}, \bibinfo{person}{Vanessa Di~Tano}, \bibinfo{person}{Chris Wright}, \bibinfo{person}{Annas Al-Sharea}, \bibinfo{person}{Joshua~G Kovoor}, \bibinfo{person}{David Gonzalez-Chica}, {and} \bibinfo{person}{Mathew~O Jacob}.} \bibinfo{year}{2026}\natexlab{}.
\newblock \showarticletitle{Artificial Intelligence-Enhanced Dictation Improves Efficiency in a Regional General Surgery Clinic: A Quasi-Experimental Pilot Study}.
\newblock \bibinfo{journal}{\emph{ANZ Journal of Surgery}} (\bibinfo{year}{2026}).
\newblock


\bibitem[Castro et~al\mbox{.}(2026)]%
        {castro2026psychiatric}
\bibfield{author}{\bibinfo{person}{Victor~M Castro}, \bibinfo{person}{Thomas~H McCoy}, \bibinfo{person}{Pilar Verhaak}, \bibinfo{person}{Anudeepa Ramachandiran}, {and} \bibinfo{person}{Roy~H Perlis}.} \bibinfo{year}{2026}\natexlab{}.
\newblock \showarticletitle{Psychiatric Documentation and Management in Primary Care With Artificial Intelligence Scribe Use}.
\newblock \bibinfo{journal}{\emph{JAMA psychiatry}} (\bibinfo{year}{2026}).
\newblock


\bibitem[Chowdhury et~al\mbox{.}(2026)]%
        {chowdhury2026comparing}
\bibfield{author}{\bibinfo{person}{Anand Chowdhury}, \bibinfo{person}{Michele Casey}, \bibinfo{person}{Jonathan Wilson}, \bibinfo{person}{Kathryn~I Pollak}, \bibinfo{person}{Benjamin~A Goldstein}, \bibinfo{person}{Armando Bedoya}, {and} \bibinfo{person}{Eric~G Poon}.} \bibinfo{year}{2026}\natexlab{}.
\newblock \showarticletitle{Comparing ambient scribes: a randomized crossover clinical trial addressing ambient scribe technologies' impact on physician burnout}.
\newblock \bibinfo{journal}{\emph{Journal of the American Medical Informatics Association}} \bibinfo{volume}{33}, \bibinfo{number}{5} (\bibinfo{year}{2026}), \bibinfo{pages}{990--999}.
\newblock


\bibitem[Downing et~al\mbox{.}(2018)]%
        {downing2018physician}
\bibfield{author}{\bibinfo{person}{N~Lance Downing}, \bibinfo{person}{David~W Bates}, {and} \bibinfo{person}{Christopher~A Longhurst}.} \bibinfo{year}{2018}\natexlab{}.
\newblock Physician burnout in the electronic health record era: are we ignoring the real cause?
\newblock \bibinfo{numpages}{50--51}~pages.
\newblock


\bibitem[Duggan et~al\mbox{.}(2025)]%
        {duggan2025clinician}
\bibfield{author}{\bibinfo{person}{Matthew~J Duggan}, \bibinfo{person}{Julietta Gervase}, \bibinfo{person}{Anna Schoenbaum}, \bibinfo{person}{William Hanson}, \bibinfo{person}{John~T Howell~III}, \bibinfo{person}{Michael Sheinberg}, {and} \bibinfo{person}{Kevin~B Johnson}.} \bibinfo{year}{2025}\natexlab{}.
\newblock \showarticletitle{Clinician experiences with ambient scribe technology to assist with documentation burden and efficiency}.
\newblock \bibinfo{journal}{\emph{JAMA Network Open}} \bibinfo{volume}{8}, \bibinfo{number}{2} (\bibinfo{year}{2025}), \bibinfo{pages}{e2460637}.
\newblock


\bibitem[Dutta et~al\mbox{.}(2026)]%
        {dutta2026medical}
\bibfield{author}{\bibinfo{person}{Sayon Dutta}, \bibinfo{person}{Jacqueline Guan-Ting You}, \bibinfo{person}{Lisette Dunham}, \bibinfo{person}{Rebecca Cash}, \bibinfo{person}{Melissa Meeker}, \bibinfo{person}{Benjamin~A White}, {and} \bibinfo{person}{Joshua~W Joseph}.} \bibinfo{year}{2026}\natexlab{}.
\newblock \showarticletitle{Medical Scribe and Ambient Artificial Intelligence Impact on Emergency Physician Documentation Burden and Clinical Productivity}.
\newblock \bibinfo{journal}{\emph{Annals of Emergency Medicine}} (\bibinfo{year}{2026}).
\newblock


\bibitem[{Elion}(2026)]%
        {elion2025scribes}
\bibfield{author}{\bibinfo{person}{{Elion}}.} \bibinfo{year}{2026}\natexlab{}.
\newblock {AI} Ambient Scribes.
\newblock \bibinfo{howpublished}{\url{https://elion.health/categories/ai-ambient-scribes/products}}.
\newblock
\newblock
\shownote{Accessed July 2026}.


\bibitem[Evans et~al\mbox{.}(2025)]%
        {evans2025impact}
\bibfield{author}{\bibinfo{person}{Kerrie Evans}, \bibinfo{person}{Amy Papinniemi}, \bibinfo{person}{Bernd Ploderer}, \bibinfo{person}{Vaughan Nicholson}, \bibinfo{person}{Tom Hindhaugh}, \bibinfo{person}{Viana Vuvan}, \bibinfo{person}{Nicholas Cowley}, \bibinfo{person}{Amina Tariq}, {and} \bibinfo{person}{Hayley Thomson}.} \bibinfo{year}{2025}\natexlab{}.
\newblock \showarticletitle{Impact of using an AI scribe on clinical documentation and clinician-patient interactions in allied health private practice: perspectives of clinicians and patients}.
\newblock \bibinfo{journal}{\emph{Musculoskeletal Science and Practice}}  \bibinfo{volume}{78} (\bibinfo{year}{2025}), \bibinfo{pages}{103333}.
\newblock


\bibitem[Gardner et~al\mbox{.}(2019)]%
        {gardner2019physician}
\bibfield{author}{\bibinfo{person}{Rebekah~L Gardner}, \bibinfo{person}{Emily Cooper}, \bibinfo{person}{Jacqueline Haskell}, \bibinfo{person}{Daniel~A Harris}, \bibinfo{person}{Sara Poplau}, \bibinfo{person}{Philip~J Kroth}, {and} \bibinfo{person}{Mark Linzer}.} \bibinfo{year}{2019}\natexlab{}.
\newblock \showarticletitle{Physician stress and burnout: the impact of health information technology}.
\newblock \bibinfo{journal}{\emph{Journal of the American Medical Informatics Association}} \bibinfo{volume}{26}, \bibinfo{number}{2} (\bibinfo{year}{2019}), \bibinfo{pages}{106--114}.
\newblock


\bibitem[Guo et~al\mbox{.}(2025)]%
        {guo2025ambient}
\bibfield{author}{\bibinfo{person}{Yawen Guo}, \bibinfo{person}{DI Hu}, \bibinfo{person}{Jiayuan Wang}, \bibinfo{person}{Kai Zheng}, \bibinfo{person}{Danielle Perret}, \bibinfo{person}{Deepti Pandita}, {and} \bibinfo{person}{Steven Tam}.} \bibinfo{year}{2025}\natexlab{}.
\newblock \showarticletitle{Ambient listening in clinical practice: evaluating EPIC signal data before and after implementation and its impact on physician workload}.
\newblock \bibinfo{journal}{\emph{arXiv preprint arXiv:2504.13879}} (\bibinfo{year}{2025}).
\newblock


\bibitem[Ha et~al\mbox{.}(2025)]%
        {ha2025evaluating}
\bibfield{author}{\bibinfo{person}{Emily Ha}, \bibinfo{person}{Isabelle Choon-Kon-Yune}, \bibinfo{person}{LaShawn Murray}, \bibinfo{person}{Siying Luan}, \bibinfo{person}{Enid Montague}, \bibinfo{person}{Onil Bhattacharyya}, {and} \bibinfo{person}{Payal Agarwal}.} \bibinfo{year}{2025}\natexlab{}.
\newblock \showarticletitle{Evaluating the usability, technical performance, and accuracy of artificial intelligence scribes for primary care: competitive analysis}.
\newblock \bibinfo{journal}{\emph{JMIR Human Factors}} \bibinfo{volume}{12}, \bibinfo{number}{1} (\bibinfo{year}{2025}), \bibinfo{pages}{e71434}.
\newblock


\bibitem[Haberle et~al\mbox{.}(2024)]%
        {haberle2024impact}
\bibfield{author}{\bibinfo{person}{Tyler Haberle}, \bibinfo{person}{Courtney Cleveland}, \bibinfo{person}{Greg~L Snow}, \bibinfo{person}{Chris Barber}, \bibinfo{person}{Nikki Stookey}, \bibinfo{person}{Cari Thornock}, \bibinfo{person}{Laurie Younger}, \bibinfo{person}{Buzzy Mullahkhel}, {and} \bibinfo{person}{Diego Ize-Ludlow}.} \bibinfo{year}{2024}\natexlab{}.
\newblock \showarticletitle{The impact of nuance DAX ambient listening AI documentation: a cohort study}.
\newblock \bibinfo{journal}{\emph{Journal of the American Medical Informatics Association}} \bibinfo{volume}{31}, \bibinfo{number}{4} (\bibinfo{year}{2024}), \bibinfo{pages}{975--979}.
\newblock


\bibitem[Halamka and Tripathi(2017)]%
        {halamka2017hitech}
\bibfield{author}{\bibinfo{person}{John~D Halamka} {and} \bibinfo{person}{Micky Tripathi}.} \bibinfo{year}{2017}\natexlab{}.
\newblock \showarticletitle{The HITECH era in retrospect}.
\newblock \bibinfo{journal}{\emph{New England Journal of Medicine}} \bibinfo{volume}{377}, \bibinfo{number}{10} (\bibinfo{year}{2017}), \bibinfo{pages}{907--909}.
\newblock


\bibitem[Harvey et~al\mbox{.}(2026)]%
        {harvey2026ambient}
\bibfield{author}{\bibinfo{person}{Cameron~J Harvey}, \bibinfo{person}{Josiah Morita}, \bibinfo{person}{William Huynh}, \bibinfo{person}{Russell~K Woo}, {and} \bibinfo{person}{Jerome~P Lee}.} \bibinfo{year}{2026}\natexlab{}.
\newblock \showarticletitle{Ambient AI Scribe Implementation in an Ambulatory Setting in a Single Medical Group: Prospective Study}.
\newblock \bibinfo{journal}{\emph{JMIR Medical Informatics}} \bibinfo{volume}{14}, \bibinfo{number}{1} (\bibinfo{year}{2026}), \bibinfo{pages}{e84104}.
\newblock


\bibitem[Harvey et~al\mbox{.}(2025)]%
        {harvey2025ambient}
\bibfield{author}{\bibinfo{person}{Cameron~J Harvey}, \bibinfo{person}{Victor Wong}, \bibinfo{person}{William Huynh}, \bibinfo{person}{Jerome~P Lee}, {and} \bibinfo{person}{Russell~K Woo}.} \bibinfo{year}{2025}\natexlab{}.
\newblock \showarticletitle{Ambient AI-assisted clinical documentation in surgical outpatient care: a preliminary study of usability, workflow, and patient experience}.
\newblock \bibinfo{journal}{\emph{World Journal of Pediatric Surgery}} \bibinfo{volume}{8}, \bibinfo{number}{5} (\bibinfo{year}{2025}), \bibinfo{pages}{e001073}.
\newblock


\bibitem[Holmgren et~al\mbox{.}(2026)]%
        {holmgren2026ambient}
\bibfield{author}{\bibinfo{person}{A~Jay Holmgren}, \bibinfo{person}{Cynthia~L Fenton}, \bibinfo{person}{Robert Thombley}, \bibinfo{person}{Hossein Soleimani}, \bibinfo{person}{Rhiannon Croci}, \bibinfo{person}{Orianna DeMasi}, \bibinfo{person}{Maria~E Byron}, \bibinfo{person}{Sara~G Murray}, \bibinfo{person}{Julia~R Adler-Milstein}, {and} \bibinfo{person}{Jinoos Yazdany}.} \bibinfo{year}{2026}\natexlab{}.
\newblock \showarticletitle{Ambient artificial intelligence scribes and physician financial productivity}.
\newblock \bibinfo{journal}{\emph{JAMA Network Open}} \bibinfo{volume}{9}, \bibinfo{number}{1} (\bibinfo{year}{2026}), \bibinfo{pages}{e2553233}.
\newblock


\bibitem[Husa et~al\mbox{.}(2026)]%
        {husa2026ambient}
\bibfield{author}{\bibinfo{person}{Robyn~A Husa}, \bibinfo{person}{John Haggerty}, \bibinfo{person}{Andrew~W Nute}, \bibinfo{person}{Julie Levine}, \bibinfo{person}{Kevin Love}, \bibinfo{person}{Xochitl Martinez}, {and} \bibinfo{person}{Canada Parrish}.} \bibinfo{year}{2026}\natexlab{}.
\newblock \showarticletitle{Ambient Artificial Intelligence Use and Clinician Documentation Burden, Productivity, and Efficiency}.
\newblock \bibinfo{journal}{\emph{JAMA Network Open}} \bibinfo{volume}{9}, \bibinfo{number}{5} (\bibinfo{year}{2026}), \bibinfo{pages}{e2615762}.
\newblock


\bibitem[Kanaparthy et~al\mbox{.}(2025)]%
        {kanaparthy2025real}
\bibfield{author}{\bibinfo{person}{Naga~Sasidhar Kanaparthy}, \bibinfo{person}{Yenny Villuendas-Rey}, \bibinfo{person}{Tolulope Bakare}, \bibinfo{person}{Zihan Diao}, \bibinfo{person}{Mark Iscoe}, \bibinfo{person}{Andrew Loza}, \bibinfo{person}{Donald Wright}, \bibinfo{person}{Conrad Safranek}, \bibinfo{person}{Isaac~V Faustino}, \bibinfo{person}{Alexandria Brackett}, {et~al\mbox{.}}} \bibinfo{year}{2025}\natexlab{}.
\newblock \showarticletitle{Real-world evidence synthesis of digital scribes using ambient listening and generative artificial intelligence for clinician documentation workflows: rapid review}.
\newblock \bibinfo{journal}{\emph{Jmir Ai}}  \bibinfo{volume}{4} (\bibinfo{year}{2025}), \bibinfo{pages}{e76743}.
\newblock


\bibitem[Kapoor et~al\mbox{.}(2026)]%
        {kapoor2026improving}
\bibfield{author}{\bibinfo{person}{Vishal Kapoor}, \bibinfo{person}{Catherine Skellern}, \bibinfo{person}{Pauline McGrath}, \bibinfo{person}{Stephanie Brosnan}, \bibinfo{person}{Yonatan Gofer}, {and} \bibinfo{person}{Linda Thorburn}.} \bibinfo{year}{2026}\natexlab{}.
\newblock \showarticletitle{Improving Clinical Efficiency Using Artificial Intelligence Scribe for Documentation (ICE-AID Study): A Retrospective-Prospective Cohort Study at a Quaternary Children's Hospital and Health Service}.
\newblock \bibinfo{journal}{\emph{Journal of Paediatrics and Child Health}} (\bibinfo{year}{2026}).
\newblock


\bibitem[Lai et~al\mbox{.}(2023)]%
        {lai2023towards}
\bibfield{author}{\bibinfo{person}{Vivian Lai}, \bibinfo{person}{Chacha Chen}, \bibinfo{person}{Alison Smith-Renner}, \bibinfo{person}{Q~Vera Liao}, {and} \bibinfo{person}{Chenhao Tan}.} \bibinfo{year}{2023}\natexlab{}.
\newblock \showarticletitle{Towards a science of human-AI decision making: An overview of design space in empirical human-subject studies}. In \bibinfo{booktitle}{\emph{Proceedings of the 2023 ACM conference on fairness, accountability, and transparency}}. \bibinfo{pages}{1369--1385}.
\newblock


\bibitem[Lawrence et~al\mbox{.}(2026)]%
        {lawrence2026implementing}
\bibfield{author}{\bibinfo{person}{Katharine Lawrence}, \bibinfo{person}{Conner Polet}, \bibinfo{person}{Kiran Malhotra}, \bibinfo{person}{Vasudev Kuram}, {and} \bibinfo{person}{Sarah Sharif}.} \bibinfo{year}{2026}\natexlab{}.
\newblock \showarticletitle{Implementing artificial intelligence-enabled ambient documentation technology for ambulatory clinicians: an innovation evaluation}.
\newblock \bibinfo{journal}{\emph{Journal of General Internal Medicine}} (\bibinfo{year}{2026}), \bibinfo{pages}{1--4}.
\newblock


\bibitem[Lukac et~al\mbox{.}(2025)]%
        {lukac2025ambient}
\bibfield{author}{\bibinfo{person}{Paul~J Lukac}, \bibinfo{person}{William Turner}, \bibinfo{person}{Sitaram Vangala}, \bibinfo{person}{Aaron~T Chin}, \bibinfo{person}{Joshua Khalili}, \bibinfo{person}{Ya-Chen~Tina Shih}, \bibinfo{person}{Catherine Sarkisian}, \bibinfo{person}{Eric~M Cheng}, {and} \bibinfo{person}{John~N Mafi}.} \bibinfo{year}{2025}\natexlab{}.
\newblock \showarticletitle{Ambient AI scribes in clinical practice: a randomized trial}.
\newblock \bibinfo{journal}{\emph{NEJM AI}} \bibinfo{volume}{2}, \bibinfo{number}{12} (\bibinfo{year}{2025}), \bibinfo{pages}{AIoa2501000}.
\newblock


\bibitem[Marquis et~al\mbox{.}(2026)]%
        {marquis2026ai}
\bibfield{author}{\bibinfo{person}{Taylor Marquis}, \bibinfo{person}{Matthew Kopp}, \bibinfo{person}{Jared~S Anderson}, \bibinfo{person}{Anthony~M Napoli}, \bibinfo{person}{Linda~L Brown}, {and} \bibinfo{person}{Yosef Berlyand}.} \bibinfo{year}{2026}\natexlab{}.
\newblock \showarticletitle{AI-powered ambient scribe technology experiences among emergency physicians: cross-sectional, mixed methods pilot survey study}.
\newblock \bibinfo{journal}{\emph{JMIR Formative Research}}  \bibinfo{volume}{10} (\bibinfo{year}{2026}), \bibinfo{pages}{e80401}.
\newblock


\bibitem[McCrudden et~al\mbox{.}(2026)]%
        {mccrudden2026ai}
\bibfield{author}{\bibinfo{person}{Kaitlin~E McCrudden}, \bibinfo{person}{Mackenzie~S Swirbul}, \bibinfo{person}{Emily~E Peake}, \bibinfo{person}{Michael~J Rodio}, {and} \bibinfo{person}{Aarthi Padmanabhan}.} \bibinfo{year}{2026}\natexlab{}.
\newblock \showarticletitle{AI-powered documentation for mental health providers: retrospective observational mixed methods study}.
\newblock \bibinfo{journal}{\emph{JMIR Formative Research}}  \bibinfo{volume}{10} (\bibinfo{year}{2026}), \bibinfo{pages}{e84628}.
\newblock


\bibitem[Memon et~al\mbox{.}(2026)]%
        {memon2026performance}
\bibfield{author}{\bibinfo{person}{Salim Memon}, \bibinfo{person}{Adam Brand}, \bibinfo{person}{Bianca Taylor}, \bibinfo{person}{Adelaide Michael}, {and} \bibinfo{person}{Rachael Smithson}.} \bibinfo{year}{2026}\natexlab{}.
\newblock \showarticletitle{Performance, acceptability, and impact of ambient listening scribe technology in an outpatient context: a mixed methods trial evaluation}.
\newblock \bibinfo{journal}{\emph{BMC Health Services Research}} (\bibinfo{year}{2026}).
\newblock


\bibitem[Moura et~al\mbox{.}(2026)]%
        {moura2026hybrid}
\bibfield{author}{\bibinfo{person}{Lidia~MVR Moura}, \bibinfo{person}{Rebecca~G Mishuris}, \bibinfo{person}{Joshua~P Metlay}, \bibinfo{person}{Mamoon Habib}, \bibinfo{person}{David~Y Ting}, \bibinfo{person}{Katherine~L Gallagher}, \bibinfo{person}{Suzanne Brodney}, \bibinfo{person}{Meghan~L Rieu-Werden}, {and} \bibinfo{person}{Jennifer~S Haas}.} \bibinfo{year}{2026}\natexlab{}.
\newblock \showarticletitle{Hybrid ambient clinical documentation and physician performance: work outside of work, documentation delay, and financial productivity}.
\newblock \bibinfo{journal}{\emph{Journal of general internal medicine}} \bibinfo{volume}{41}, \bibinfo{number}{5} (\bibinfo{year}{2026}), \bibinfo{pages}{1294--1303}.
\newblock


\bibitem[{National Institute of Standards and Technology}(2023)]%
        {nist_ai_rmf}
\bibfield{author}{\bibinfo{person}{{National Institute of Standards and Technology}}.} \bibinfo{year}{2023}\natexlab{}.
\newblock Artificial Intelligence Risk Management Framework ({AI} {RMF} 1.0).
\newblock \bibinfo{howpublished}{\url{https://nvlpubs.nist.gov/nistpubs/ai/NIST.AI.100-1.pdf}}.
\newblock
\newblock
\shownote{NIST AI 100-1. Accessed: 2026-08-01}.


\bibitem[Nguyen et~al\mbox{.}(2023)]%
        {nguyen2023implementing}
\bibfield{author}{\bibinfo{person}{Oliver~T Nguyen}, \bibinfo{person}{Kea Turner}, \bibinfo{person}{Dannelle Charles}, \bibinfo{person}{Olivia Sprow}, \bibinfo{person}{Randa Perkins}, \bibinfo{person}{Young-Rock Hong}, \bibinfo{person}{Jessica~Y Islam}, \bibinfo{person}{Neel Khanna}, \bibinfo{person}{Amir Alishahi~Tabriz}, \bibinfo{person}{Julie Hallanger-Johnson}, {et~al\mbox{.}}} \bibinfo{year}{2023}\natexlab{}.
\newblock \showarticletitle{Implementing digital scribes to reduce electronic health record documentation burden among cancer care clinicians: a mixed-methods pilot study}.
\newblock \bibinfo{journal}{\emph{JCO Clinical Cancer Informatics}}  \bibinfo{volume}{7} (\bibinfo{year}{2023}), \bibinfo{pages}{e2200166}.
\newblock


\bibitem[North et~al\mbox{.}(2025)]%
        {north2025ambient}
\bibfield{author}{\bibinfo{person}{Frederick North}, \bibinfo{person}{Marc~R Matthews}, \bibinfo{person}{Asif Iqbal}, \bibinfo{person}{Jason~A Post}, {and} \bibinfo{person}{Jon~O Ebbert}.} \bibinfo{year}{2025}\natexlab{}.
\newblock \showarticletitle{Ambient listening implementation in primary care and changes in electronic health record documentation metrics: Pre--post study of an ambient listening tool}.
\newblock \bibinfo{journal}{\emph{Digital Health}}  \bibinfo{volume}{11} (\bibinfo{year}{2025}), \bibinfo{pages}{20552076251403211}.
\newblock


\bibitem[Olson et~al\mbox{.}(2025)]%
        {olson2025use}
\bibfield{author}{\bibinfo{person}{Kristine~D Olson}, \bibinfo{person}{Daniella Meeker}, \bibinfo{person}{Matt Troup}, \bibinfo{person}{Timothy~D Barker}, \bibinfo{person}{Vinh~H Nguyen}, \bibinfo{person}{Jennifer~B Manders}, \bibinfo{person}{Cheryl~D Stults}, \bibinfo{person}{Veena~G Jones}, \bibinfo{person}{Sachin~D Shah}, \bibinfo{person}{Tina Shah}, {et~al\mbox{.}}} \bibinfo{year}{2025}\natexlab{}.
\newblock \showarticletitle{Use of ambient AI scribes to reduce administrative burden and professional burnout}.
\newblock \bibinfo{journal}{\emph{JAMA Network Open}} \bibinfo{volume}{8}, \bibinfo{number}{10} (\bibinfo{year}{2025}), \bibinfo{pages}{e2534976}.
\newblock


\bibitem[Page et~al\mbox{.}(2021)]%
        {page2021prisma}
\bibfield{author}{\bibinfo{person}{Matthew~J Page}, \bibinfo{person}{Joanne~E McKenzie}, \bibinfo{person}{Patrick~M Bossuyt}, \bibinfo{person}{Isabelle Boutron}, \bibinfo{person}{Tammy~C Hoffmann}, \bibinfo{person}{Cynthia~D Mulrow}, \bibinfo{person}{Larissa Shamseer}, \bibinfo{person}{Jennifer~M Tetzlaff}, \bibinfo{person}{Elie~A Akl}, \bibinfo{person}{Sue~E Brennan}, {et~al\mbox{.}}} \bibinfo{year}{2021}\natexlab{}.
\newblock \showarticletitle{The PRISMA 2020 statement: an updated guideline for reporting systematic reviews}.
\newblock \bibinfo{journal}{\emph{bmj}}  \bibinfo{volume}{372} (\bibinfo{year}{2021}).
\newblock


\bibitem[Patel et~al\mbox{.}(2026)]%
        {patel2026effect}
\bibfield{author}{\bibinfo{person}{Kajal~N Patel}, \bibinfo{person}{Mary~M Stout}, \bibinfo{person}{Mary~A Solis}, \bibinfo{person}{Nasrin Riyahi}, \bibinfo{person}{Christopher Holland}, \bibinfo{person}{Pamela~D Vohra-Khullar}, \bibinfo{person}{Miranda~A Moore}, {and} \bibinfo{person}{Reema~H Dbouk}.} \bibinfo{year}{2026}\natexlab{}.
\newblock \showarticletitle{The Effect of Ambient Listening Technology on the Patient Experience}.
\newblock \bibinfo{journal}{\emph{Applied Clinical Informatics}} \bibinfo{volume}{17}, \bibinfo{number}{01} (\bibinfo{year}{2026}), \bibinfo{pages}{82--88}.
\newblock


\bibitem[Pearlman et~al\mbox{.}(2025)]%
        {pearlman2025use}
\bibfield{author}{\bibinfo{person}{Kevin Pearlman}, \bibinfo{person}{Wen Wan}, \bibinfo{person}{Sachin Shah}, {and} \bibinfo{person}{Neda Laiteerapong}.} \bibinfo{year}{2025}\natexlab{}.
\newblock \showarticletitle{Use of an AI scribe and electronic health record efficiency}.
\newblock \bibinfo{journal}{\emph{JAMA Network Open}} \bibinfo{volume}{8}, \bibinfo{number}{10} (\bibinfo{year}{2025}), \bibinfo{pages}{e2537000}.
\newblock


\bibitem[Pelletier et~al\mbox{.}(2025)]%
        {pelletier2025effect}
\bibfield{author}{\bibinfo{person}{Jonathan~H Pelletier}, \bibinfo{person}{Kevin Watson}, \bibinfo{person}{Jenny Michel}, \bibinfo{person}{Robert McGregor}, {and} \bibinfo{person}{Sarah~Z Rush}.} \bibinfo{year}{2025}\natexlab{}.
\newblock \showarticletitle{Effect of a generative artificial intelligence digital scribe on pediatric provider documentation time, cognitive burden, and burnout}.
\newblock \bibinfo{journal}{\emph{JAMIA open}} \bibinfo{volume}{8}, \bibinfo{number}{4} (\bibinfo{year}{2025}), \bibinfo{pages}{ooaf068}.
\newblock


\bibitem[Preiksaitis et~al\mbox{.}(2026a)]%
        {preiksaitis2026ambientAI}
\bibfield{author}{\bibinfo{person}{Carl Preiksaitis}, \bibinfo{person}{Al'ai Alvarez}, \bibinfo{person}{Maia Winkel}, \bibinfo{person}{Mia Karamatsu}, \bibinfo{person}{Ian Brown}, \bibinfo{person}{Neetha Sama}, \bibinfo{person}{Luke Morris}, \bibinfo{person}{Jae-yeon Lee}, \bibinfo{person}{Allie Gubbels}, \bibinfo{person}{Eileen Wahl}, {et~al\mbox{.}}} \bibinfo{year}{2026}\natexlab{a}.
\newblock \showarticletitle{Ambient AI Scribes and Emergency Department Documentation Burden: Retrospective Cohort Study}.
\newblock \bibinfo{journal}{\emph{JMIR AI}}  \bibinfo{volume}{5} (\bibinfo{year}{2026}), \bibinfo{pages}{e92193}.
\newblock


\bibitem[Preiksaitis et~al\mbox{.}(2026b)]%
        {preiksaitis2026ambient}
\bibfield{author}{\bibinfo{person}{Carl Preiksaitis}, \bibinfo{person}{AlÕai Alvarez}, \bibinfo{person}{Maia Winkel}, \bibinfo{person}{Mia Karamatsu}, \bibinfo{person}{Ian Brown}, \bibinfo{person}{Neetha Sama}, \bibinfo{person}{Luke Morris}, \bibinfo{person}{Jae-yeon Lee}, \bibinfo{person}{Allie Gubbels}, \bibinfo{person}{Eileen Wahl}, {et~al\mbox{.}}} \bibinfo{year}{2026}\natexlab{b}.
\newblock \showarticletitle{Ambient artificial intelligence scribe adoption and documentation time in the emergency department}.
\newblock \bibinfo{journal}{\emph{Annals of Emergency Medicine}} (\bibinfo{year}{2026}).
\newblock


\bibitem[Rabbani et~al\mbox{.}(2025)]%
        {rabbani2025ambient}
\bibfield{author}{\bibinfo{person}{Naveed Rabbani}, \bibinfo{person}{Mondira Ray}, \bibinfo{person}{Eleanor Verhagen}, \bibinfo{person}{Jonathan Hatoun}, \bibinfo{person}{Laura~Burckett Patane}, {and} \bibinfo{person}{Louis Vernacchio}.} \bibinfo{year}{2025}\natexlab{}.
\newblock \showarticletitle{Ambient Artificial Intelligence Scribes in Pediatric Primary Care: A Mixed Methods Study}.
\newblock \bibinfo{journal}{\emph{Applied Clinical Informatics}} \bibinfo{volume}{16}, \bibinfo{number}{05} (\bibinfo{year}{2025}), \bibinfo{pages}{1578--1587}.
\newblock


\bibitem[Redhead(2009)]%
        {act2009health}
\bibfield{author}{\bibinfo{person}{C.~Stephen Redhead}.} \bibinfo{year}{2009}\natexlab{}.
\newblock Health information technology for economic and clinical health.
\newblock \bibinfo{howpublished}{\url{https://www.congress.gov/crs-product/R40161}}.
\newblock


\bibitem[Reverberi et~al\mbox{.}(2022)]%
        {reverberi2022experimental}
\bibfield{author}{\bibinfo{person}{Carlo Reverberi}, \bibinfo{person}{Tommaso Rigon}, \bibinfo{person}{Aldo Solari}, \bibinfo{person}{Cesare Hassan}, \bibinfo{person}{Paolo Cherubini}, {and} \bibinfo{person}{Andrea Cherubini}.} \bibinfo{year}{2022}\natexlab{}.
\newblock \showarticletitle{Experimental evidence of effective human--AI collaboration in medical decision-making}.
\newblock \bibinfo{journal}{\emph{Scientific reports}} \bibinfo{volume}{12}, \bibinfo{number}{1} (\bibinfo{year}{2022}), \bibinfo{pages}{14952}.
\newblock


\bibitem[Rotenstein et~al\mbox{.}(2026)]%
        {rotenstein2026changes}
\bibfield{author}{\bibinfo{person}{Lisa~S Rotenstein}, \bibinfo{person}{A~Jay Holmgren}, \bibinfo{person}{Robert Thombley}, \bibinfo{person}{Aditi Sriram}, \bibinfo{person}{Reema~H Dbouk}, \bibinfo{person}{Melissa Jost}, \bibinfo{person}{Debbie Aizenberg}, \bibinfo{person}{Scott MacDonald}, \bibinfo{person}{Naga Kanaparthy}, \bibinfo{person}{Brian Williams}, {et~al\mbox{.}}} \bibinfo{year}{2026}\natexlab{}.
\newblock \showarticletitle{Changes in clinician time expenditure and visit quantity with adoption of artificial intelligence--powered scribes: a multisite study}.
\newblock \bibinfo{journal}{\emph{JAMA}} \bibinfo{volume}{335}, \bibinfo{number}{16} (\bibinfo{year}{2026}), \bibinfo{pages}{1408--1417}.
\newblock


\bibitem[Rui et~al\mbox{.}(2025)]%
        {rui2025qualitative}
\bibfield{author}{\bibinfo{person}{Angela Rui}, \bibinfo{person}{Gabrielle Choo-Kang}, \bibinfo{person}{Ivana Salmikova}, \bibinfo{person}{Christine Iannaccone}, \bibinfo{person}{Jacqueline You}, \bibinfo{person}{Christopher Holland}, \bibinfo{person}{Bryan Blanchette}, \bibinfo{person}{Reema Dbouk}, \bibinfo{person}{Rebecca Mishuris}, {and} \bibinfo{person}{Lisa Rotenstein}.} \bibinfo{year}{2025}\natexlab{}.
\newblock \showarticletitle{A Qualitative Study on Clinician Perceptions of Ambient Scribe Documentation}. In \bibinfo{booktitle}{\emph{2025 Annual Research Meeting}}. AcademyHealth.
\newblock


\bibitem[Rule et~al\mbox{.}(2021)]%
        {rule2021length}
\bibfield{author}{\bibinfo{person}{Adam Rule}, \bibinfo{person}{Steven Bedrick}, \bibinfo{person}{Michael~F Chiang}, {and} \bibinfo{person}{Michelle~R Hribar}.} \bibinfo{year}{2021}\natexlab{}.
\newblock \showarticletitle{Length and redundancy of outpatient progress notes across a decade at an academic medical center}.
\newblock \bibinfo{journal}{\emph{JAMA Network Open}} \bibinfo{volume}{4}, \bibinfo{number}{7} (\bibinfo{year}{2021}), \bibinfo{pages}{e2115334}.
\newblock


\bibitem[Sanmark et~al\mbox{.}(2026)]%
        {sanmark2026evaluation}
\bibfield{author}{\bibinfo{person}{Enni Sanmark}, \bibinfo{person}{Ville Vartiainen}, \bibinfo{person}{Johan Sanmark}, \bibinfo{person}{Katarina Wettin}, \bibinfo{person}{Lukas Saari}, {and} \bibinfo{person}{Artin Entezarjou}.} \bibinfo{year}{2026}\natexlab{}.
\newblock \showarticletitle{Evaluation of an AI Medical Scribe After 236,153 Notes Generated Across Care Levels in a European Health System: Mixed Methods Retrospective Observational Study}.
\newblock \bibinfo{journal}{\emph{JMIR medical informatics}} \bibinfo{volume}{14}, \bibinfo{number}{1} (\bibinfo{year}{2026}), \bibinfo{pages}{e90052}.
\newblock


\bibitem[Sasseville et~al\mbox{.}(2025)]%
        {sasseville2025impact}
\bibfield{author}{\bibinfo{person}{Maxime Sasseville}, \bibinfo{person}{Farzaneh Yousefi}, \bibinfo{person}{Steven Ouellet}, \bibinfo{person}{Florian Naye}, \bibinfo{person}{Th{\'e}o Stefan}, \bibinfo{person}{Val{\'e}rie Carnovale}, \bibinfo{person}{Fr{\'e}d{\'e}ric Bergeron}, \bibinfo{person}{Linda Ling}, \bibinfo{person}{Bobby Gheorghiu}, \bibinfo{person}{Simon Hagens}, {et~al\mbox{.}}} \bibinfo{year}{2025}\natexlab{}.
\newblock \showarticletitle{The impact of AI scribes on streamlining clinical documentation: a systematic review}. In \bibinfo{booktitle}{\emph{Healthcare}}, Vol.~\bibinfo{volume}{13}. MDPI, \bibinfo{pages}{1447}.
\newblock


\bibitem[Schneider et~al\mbox{.}(2026)]%
        {schneider2026impact}
\bibfield{author}{\bibinfo{person}{Kathryn~R Schneider}, \bibinfo{person}{Hillary~E Swann-Thomsen}, \bibinfo{person}{Terry~G Ribbens}, \bibinfo{person}{Lucas~A Bahnmaier}, \bibinfo{person}{Trevor Satterfield}, \bibinfo{person}{Reme Pullicar}, {and} \bibinfo{person}{Neeraj Soni}.} \bibinfo{year}{2026}\natexlab{}.
\newblock \showarticletitle{The impact of artificial intelligence scribes on physician and advanced practice provider cognitive load and well-being}.
\newblock \bibinfo{journal}{\emph{Journal of the American Medical Informatics Association}} \bibinfo{volume}{33}, \bibinfo{number}{5} (\bibinfo{year}{2026}), \bibinfo{pages}{973--978}.
\newblock


\bibitem[Shah et~al\mbox{.}(2025)]%
        {shah2025ambient}
\bibfield{author}{\bibinfo{person}{Shreya~J Shah}, \bibinfo{person}{Anna Devon-Sand}, \bibinfo{person}{Stephen~P Ma}, \bibinfo{person}{Yejin Jeong}, \bibinfo{person}{Trevor Crowell}, \bibinfo{person}{Margaret Smith}, \bibinfo{person}{April~S Liang}, \bibinfo{person}{Clarissa Delahaie}, \bibinfo{person}{Caroline Hsia}, \bibinfo{person}{Tait Shanafelt}, {et~al\mbox{.}}} \bibinfo{year}{2025}\natexlab{}.
\newblock \showarticletitle{Ambient artificial intelligence scribes: physician burnout and perspectives on usability and documentation burden}.
\newblock \bibinfo{journal}{\emph{Journal of the American Medical Informatics Association}} \bibinfo{volume}{32}, \bibinfo{number}{2} (\bibinfo{year}{2025}), \bibinfo{pages}{375--380}.
\newblock


\bibitem[Shin et~al\mbox{.}(2025)]%
        {shin2025influence}
\bibfield{author}{\bibinfo{person}{H~Stella Shin}, \bibinfo{person}{Herb Williams}, \bibinfo{person}{Nikolay Braykov}, \bibinfo{person}{Afrin Jahan}, \bibinfo{person}{Jeremy Meller}, {and} \bibinfo{person}{Evan~W Orenstein}.} \bibinfo{year}{2025}\natexlab{}.
\newblock \showarticletitle{The influence of artificial intelligence scribes on clinician experience and efficiency among pediatric subspecialists: A rapid, randomized quality improvement trial}.
\newblock \bibinfo{journal}{\emph{Applied Clinical Informatics}} \bibinfo{volume}{16}, \bibinfo{number}{04} (\bibinfo{year}{2025}), \bibinfo{pages}{1041--1052}.
\newblock


\bibitem[Sinsky et~al\mbox{.}(2016)]%
        {sinsky2016allocation}
\bibfield{author}{\bibinfo{person}{Christine Sinsky}, \bibinfo{person}{Lacey Colligan}, \bibinfo{person}{Ling Li}, \bibinfo{person}{Mirela Prgomet}, \bibinfo{person}{Sam Reynolds}, \bibinfo{person}{Lindsey Goeders}, \bibinfo{person}{Johanna Westbrook}, \bibinfo{person}{Michael Tutty}, {and} \bibinfo{person}{George Blike}.} \bibinfo{year}{2016}\natexlab{}.
\newblock \showarticletitle{Allocation of physician time in ambulatory practice: a time and motion study in 4 specialties}.
\newblock \bibinfo{journal}{\emph{Annals of internal medicine}} \bibinfo{volume}{165}, \bibinfo{number}{11} (\bibinfo{year}{2016}), \bibinfo{pages}{753--760}.
\newblock


\bibitem[Talman et~al\mbox{.}(2026)]%
        {talman2026ambient}
\bibfield{author}{\bibinfo{person}{Molly Talman}, \bibinfo{person}{Lamia Alam}, \bibinfo{person}{Dara Mize}, \bibinfo{person}{Yaa Kumah-Crystal}, \bibinfo{person}{Carolynn Nall}, \bibinfo{person}{Allison~B McCoy}, \bibinfo{person}{Adam Wright}, \bibinfo{person}{Laura Zahn}, {and} \bibinfo{person}{Jennifer Andrews}.} \bibinfo{year}{2026}\natexlab{}.
\newblock \showarticletitle{Ambient Artificial Intelligence Scribes in Pediatric Hematology—Oncology: Early Implementation of DAX Copilot}.
\newblock \bibinfo{journal}{\emph{Applied clinical informatics}} \bibinfo{volume}{17}, \bibinfo{number}{02} (\bibinfo{year}{2026}), \bibinfo{pages}{374--379}.
\newblock


\bibitem[Tan et~al\mbox{.}(2026)]%
        {tan2026impact}
\bibfield{author}{\bibinfo{person}{Jonathan Yue~En Tan}, \bibinfo{person}{Iffat Bin~Mohammad Rafi}, \bibinfo{person}{Gerald Gui~Ren Sng}, \bibinfo{person}{Joshua Yi~Min Tung}, \bibinfo{person}{Daniel Yan~Zheng Lim}, \bibinfo{person}{Jasmine Chiat~Ling Ong}, \bibinfo{person}{Gek~Hsiang Lim}, \bibinfo{person}{Eileen Yi~Lee Lew}, \bibinfo{person}{Kuok~Wei Chia}, {and} \bibinfo{person}{Mark Kok~Hong Heng}.} \bibinfo{year}{2026}\natexlab{}.
\newblock \showarticletitle{Impact of an ambient AI scribe among clinicians and patients: real-world prospective observational time-motion study}.
\newblock \bibinfo{journal}{\emph{JMIR Medical Informatics}}  \bibinfo{volume}{14} (\bibinfo{year}{2026}), \bibinfo{pages}{e85580}.
\newblock


\bibitem[Taskforce(2025)]%
        {peterson2025adoption}
\bibfield{author}{\bibinfo{person}{Peterson Health Technology Institute~AI Taskforce}.} \bibinfo{year}{2025}\natexlab{}.
\newblock Adoption of Artificial Intelligence in Healthcare Delivery Systems: Early Applications and Impacts.
\newblock


\bibitem[Tierney et~al\mbox{.}(2025)]%
        {tierney2025ambient}
\bibfield{author}{\bibinfo{person}{Aaron~A Tierney}, \bibinfo{person}{Gregg Gayre}, \bibinfo{person}{Brian Hoberman}, \bibinfo{person}{Britt Mattern}, \bibinfo{person}{Manuel Ballesca}, \bibinfo{person}{Sarah~B Wilson~Hannay}, \bibinfo{person}{Kate Castilla}, \bibinfo{person}{Cindy~S Lau}, \bibinfo{person}{Patricia Kipnis}, \bibinfo{person}{Vincent Liu}, {et~al\mbox{.}}} \bibinfo{year}{2025}\natexlab{}.
\newblock \showarticletitle{Ambient artificial intelligence scribes: learnings after 1 year and over 2.5 million uses}.
\newblock \bibinfo{journal}{\emph{NEJM Catalyst Innovations in Care Delivery}} \bibinfo{volume}{6}, \bibinfo{number}{5} (\bibinfo{year}{2025}), \bibinfo{pages}{CAT--25}.
\newblock


\bibitem[Toussi et~al\mbox{.}(2026)]%
        {toussi2026impact}
\bibfield{author}{\bibinfo{person}{Nima Toussi}, \bibinfo{person}{Caroline Zhang}, \bibinfo{person}{Jocelyn Kang}, {and} \bibinfo{person}{Edward~J Licitra}.} \bibinfo{year}{2026}\natexlab{}.
\newblock \showarticletitle{Impact of AI Scribes on Physician Productivity and Workflow in Medical Oncology}.
\newblock \bibinfo{journal}{\emph{JCO oncology practice}} (\bibinfo{year}{2026}), \bibinfo{pages}{OP--25}.
\newblock


\bibitem[{U.S. Food and Drug Administration}(2018)]%
        {fda_samd}
\bibfield{author}{\bibinfo{person}{{U.S. Food and Drug Administration}}.} \bibinfo{year}{2018}\natexlab{}.
\newblock Software as a Medical Device ({SaMD}).
\newblock \bibinfo{howpublished}{\url{https://www.fda.gov/medical-devices/digital-health-center-excellence/software-medical-device-samd}}.
\newblock
\newblock
\shownote{Accessed: 2026-08-01}.


\bibitem[Vallabhaneni et~al\mbox{.}(2026)]%
        {vallabhaneni2026reduction}
\bibfield{author}{\bibinfo{person}{Nikhil Vallabhaneni}, \bibinfo{person}{H~Sadiyya Ingawa}, \bibinfo{person}{Brenna~E Blackburn}, {and} \bibinfo{person}{Christopher~E Pelt}.} \bibinfo{year}{2026}\natexlab{}.
\newblock \showarticletitle{Reduction of Electronic Medical Record Documentation Time and Burden Using an Artificial Intelligence Scribe Tool in an Orthopaedic Clinic}.
\newblock \bibinfo{journal}{\emph{Journal of the American Academy of Orthopaedic Surgeons}} (\bibinfo{year}{2026}).
\newblock


\bibitem[van Linschoten et~al\mbox{.}(2026)]%
        {van2026ambient}
\bibfield{author}{\bibinfo{person}{RCA van Linschoten}, \bibinfo{person}{CM van Loon}, \bibinfo{person}{L Joanknecht}, {and} \bibinfo{person}{EWMA Bischoff}.} \bibinfo{year}{2026}\natexlab{}.
\newblock \showarticletitle{Ambient scribe in general practice: a multi-perspective before-after longitudinal mixed-methods study}.
\newblock \bibinfo{journal}{\emph{npj Digital Medicine}} \bibinfo{volume}{9}, \bibinfo{number}{1} (\bibinfo{year}{2026}), \bibinfo{pages}{299}.
\newblock


\bibitem[Vasey et~al\mbox{.}(2022)]%
        {vasey2022reporting}
\bibfield{author}{\bibinfo{person}{Baptiste Vasey}, \bibinfo{person}{Myura Nagendran}, \bibinfo{person}{Bruce Campbell}, \bibinfo{person}{David~A Clifton}, \bibinfo{person}{Gary~S Collins}, \bibinfo{person}{Spiros Denaxas}, \bibinfo{person}{Alastair~K Denniston}, \bibinfo{person}{Livia Faes}, \bibinfo{person}{Bart Geerts}, \bibinfo{person}{Mudathir Ibrahim}, {et~al\mbox{.}}} \bibinfo{year}{2022}\natexlab{}.
\newblock \showarticletitle{Reporting guideline for the early-stage clinical evaluation of decision support systems driven by artificial intelligence: DECIDE-AI}.
\newblock \bibinfo{journal}{\emph{Nature medicine}} \bibinfo{volume}{28}, \bibinfo{number}{5} (\bibinfo{year}{2022}), \bibinfo{pages}{924--933}.
\newblock


\bibitem[Vereschak et~al\mbox{.}(2021)]%
        {vereschak2021evaluate}
\bibfield{author}{\bibinfo{person}{Oleksandra Vereschak}, \bibinfo{person}{Gilles Bailly}, {and} \bibinfo{person}{Baptiste Caramiaux}.} \bibinfo{year}{2021}\natexlab{}.
\newblock \showarticletitle{How to evaluate trust in AI-assisted decision making? A survey of empirical methodologies}.
\newblock \bibinfo{journal}{\emph{Proceedings of the ACM on Human-Computer Interaction}} \bibinfo{volume}{5}, \bibinfo{number}{CSCW2} (\bibinfo{year}{2021}), \bibinfo{pages}{1--39}.
\newblock


\bibitem[Waken et~al\mbox{.}(2026)]%
        {waken2026longitudinal}
\bibfield{author}{\bibinfo{person}{RJ Waken}, \bibinfo{person}{Sunny~S Lou}, \bibinfo{person}{Mackenzie Hofford}, \bibinfo{person}{Elise Eiden}, \bibinfo{person}{Carrie Burk}, \bibinfo{person}{Seunghwan Kim}, \bibinfo{person}{Jared Esker}, \bibinfo{person}{Linying Zhang}, \bibinfo{person}{Thomas~M Maddox}, \bibinfo{person}{Joanna Abraham}, {et~al\mbox{.}}} \bibinfo{year}{2026}\natexlab{}.
\newblock \showarticletitle{Longitudinal effects of ambient AI scribe use on documentation burden and financial productivity: A quasi-experimental study}.
\newblock \bibinfo{journal}{\emph{medRxiv}} (\bibinfo{year}{2026}), \bibinfo{pages}{2026--01}.
\newblock


\bibitem[Wang et~al\mbox{.}(2025)]%
        {wang2025evaluation}
\bibfield{author}{\bibinfo{person}{Haoyuan Wang}, \bibinfo{person}{Rui Yang}, \bibinfo{person}{Mahmoud Alwakeel}, \bibinfo{person}{Ankit Kayastha}, \bibinfo{person}{Anand Chowdhury}, \bibinfo{person}{Joshua~M Biro}, \bibinfo{person}{Anthony~D Sorrentino}, \bibinfo{person}{Jessica~L Handley}, \bibinfo{person}{Sarah Hantzmon}, \bibinfo{person}{Sophia Bessias}, {et~al\mbox{.}}} \bibinfo{year}{2025}\natexlab{}.
\newblock \showarticletitle{An evaluation framework for ambient digital scribing tools in clinical applications}.
\newblock \bibinfo{journal}{\emph{NPJ Digital Medicine}} \bibinfo{volume}{8}, \bibinfo{number}{1} (\bibinfo{year}{2025}), \bibinfo{pages}{358}.
\newblock


\bibitem[Wells et~al\mbox{.}(2025)]%
        {wells2025practical}
\bibfield{author}{\bibinfo{person}{Brian~J Wells}, \bibinfo{person}{Hieu~M Nguyen}, \bibinfo{person}{Andrew McWilliams}, \bibinfo{person}{Matt Pallini}, \bibinfo{person}{Amy Bovi}, \bibinfo{person}{Andrew Kuzma}, \bibinfo{person}{Justin Kramer}, \bibinfo{person}{Shih-Hsiung Chou}, \bibinfo{person}{Timothy Hetherington}, \bibinfo{person}{Patricia Corn}, {et~al\mbox{.}}} \bibinfo{year}{2025}\natexlab{}.
\newblock \showarticletitle{A practical framework for appropriate implementation and review of artificial intelligence (FAIR-AI) in healthcare}.
\newblock \bibinfo{journal}{\emph{NPJ digital medicine}} \bibinfo{volume}{8}, \bibinfo{number}{1} (\bibinfo{year}{2025}), \bibinfo{pages}{514}.
\newblock


\bibitem[{White House Office of Science and Technology Policy}(2022)]%
        {ostp_ai_bill_of_rights}
\bibfield{author}{\bibinfo{person}{{White House Office of Science and Technology Policy}}.} \bibinfo{year}{2022}\natexlab{}.
\newblock Blueprint for an {AI} Bill of Rights: Making Automated Systems Work for the American People.
\newblock \bibinfo{howpublished}{\url{https://www.govinfo.gov/app/details/GOVPUB-PREX23-PURL-gpo193638}}.
\newblock
\newblock
\shownote{Accessed: 2026-08-01}.


\bibitem[Wojda et~al\mbox{.}(2026)]%
        {wojda2026burden}
\bibfield{author}{\bibinfo{person}{Thomas Wojda}, \bibinfo{person}{Jennifer Berliner}, \bibinfo{person}{Gary~S Fischer}, \bibinfo{person}{Michael Curren}, \bibinfo{person}{Jonathan Arnold}, {and} \bibinfo{person}{Lauren Giugale}.} \bibinfo{year}{2026}\natexlab{}.
\newblock \showarticletitle{From Burden to Balance: Trends in Documentation Workload with Ambient Artificial Intelligence Scribe Adoption}.
\newblock \bibinfo{journal}{\emph{Applied Clinical Informatics}} \bibinfo{volume}{17}, \bibinfo{number}{03} (\bibinfo{year}{2026}), \bibinfo{pages}{484--492}.
\newblock


\bibitem[Wright et~al\mbox{.}(2026)]%
        {wright2026enterprise}
\bibfield{author}{\bibinfo{person}{Aileen~P Wright}, \bibinfo{person}{Carolynn~K Nall}, \bibinfo{person}{Jacob~JH Franklin}, \bibinfo{person}{Sara~N Horst}, \bibinfo{person}{Yaa~A Kumah-Crystal}, \bibinfo{person}{Adam~T Wright}, {and} \bibinfo{person}{Dara~E Mize}.} \bibinfo{year}{2026}\natexlab{}.
\newblock \showarticletitle{Enterprise-wide simultaneous deployment of ambient scribe technology: lessons learned from an academic health system}.
\newblock \bibinfo{journal}{\emph{Journal of the American Medical Informatics Association}} \bibinfo{volume}{33}, \bibinfo{number}{2} (\bibinfo{year}{2026}), \bibinfo{pages}{457--461}.
\newblock


\bibitem[Wright et~al\mbox{.}(2025)]%
        {wright2025effect}
\bibfield{author}{\bibinfo{person}{Donald~S Wright}, \bibinfo{person}{Naga~S Kanaparthy}, \bibinfo{person}{Edward~R Melnick}, \bibinfo{person}{Deborah~R Levy}, \bibinfo{person}{Stephen~J Huot}, \bibinfo{person}{Allen Hsiao}, \bibinfo{person}{Lee~H Schwamm}, {and} \bibinfo{person}{Shawn~Y Ong}.} \bibinfo{year}{2025}\natexlab{}.
\newblock \showarticletitle{The effect of ambient artificial intelligence scribes on trainee documentation burden}.
\newblock \bibinfo{journal}{\emph{Applied Clinical Informatics}} \bibinfo{volume}{16}, \bibinfo{number}{04} (\bibinfo{year}{2025}), \bibinfo{pages}{872--878}.
\newblock


\bibitem[Wu et~al\mbox{.}(2024)]%
        {wu2024evaluating}
\bibfield{author}{\bibinfo{person}{Yuxuan Wu}, \bibinfo{person}{Mingyue Wu}, \bibinfo{person}{Changyu Wang}, \bibinfo{person}{Jie Lin}, \bibinfo{person}{Jialin Liu}, {and} \bibinfo{person}{Siru Liu}.} \bibinfo{year}{2024}\natexlab{}.
\newblock \showarticletitle{Evaluating the prevalence of burnout among health care professionals related to electronic health record use: systematic review and meta-analysis}.
\newblock \bibinfo{journal}{\emph{JMIR Medical Informatics}}  \bibinfo{volume}{12} (\bibinfo{year}{2024}), \bibinfo{pages}{e54811}.
\newblock


\end{thebibliography}
\newpage
\section*{Appendix}
\appendix
%\appendixpage
\section{Search Strategy}\label{app:search}
% (paste full database search strings here)

\begin{longtable}{
    >{\RaggedRight\arraybackslash}p{0.17\textwidth}
    >{\RaggedRight\arraybackslash}p{0.75\textwidth}
}
\caption{Search strategy for all databases.}
\label{tab:search-strategy}\\

\toprule
\textbf{Database} & \textbf{Search strategy} \\
\midrule
\endfirsthead

\multicolumn{2}{l}{\textit{Table \thetable\ continued from previous page}}\\
\toprule
\textbf{Database} & \textbf{Search strategy} \\
\midrule
\endhead

\midrule
\multicolumn{2}{r}{\textit{Continued on next page}}\\
\endfoot

\bottomrule
\endlastfoot

\multicolumn{2}{l}{\textbf{Search period: November 2024--July 2026}}\\
\midrule

\textbf{Google Scholar}
&
``Ambient scribe'' OR ``Clinical scribe'' OR ``Medical scribe'' OR
``automated scribe'' OR ``virtual scribe'' OR ``Digital scribe'' OR
``AI scribe'' OR ``E scribe''
\\
\midrule

\textbf{arXiv}
&
``Ambient scribe'' OR ``Clinical scribe'' OR ``Medical scribe'' OR
``automated scribe'' OR ``virtual scribe'' OR ``Digital scribe'' OR
``AI scribe'' OR ``E scribe''
\\
\midrule

\textbf{Web of Science}
&
``Ambient scribe'' OR ``Clinical scribe'' OR ``Medical scribe'' OR
``automated scribe'' OR ``virtual scribe'' OR ``Digital scribe'' OR
``AI scribe'' OR ``E scribe''
\\
\midrule

\textbf{MEDLINE (Ovid)}
&
((Ambient OR Clinical OR Medical OR automated OR virtual OR Digital OR AI OR
``Artificial Intelligence'' OR E) adj1 scrib*).ti,ab,kf
\\
\midrule

\textbf{Embase (Embase.com)}
&
((Ambient OR Clinical OR Medical OR automated OR virtual OR Digital OR AI OR
``Artificial Intelligence'' OR E) NEAR/1 scrib*):ti,ab,kw
\\
\midrule

\textbf{CINAHL (EBSCO)}
&
TI ((Ambient OR Clinical OR Medical OR automated OR virtual OR Digital OR AI OR
``Artificial Intelligence'' OR E) N1 scrib*) OR AB ((Ambient OR Clinical OR Medical OR
automated OR virtual OR Digital OR AI OR ``Artificial Intelligence'' OR E) N1 scrib*)
\\

\end{longtable}
 
%\section{Per-Study Characteristics}\label{app:details}
%\input{details_table.tex}

 \newpage
\end{document}